\documentclass[galaxies,review,accept,pdftex,oneauthor]{Definitions/mdpi} 

\usepackage{amsmath,amssymb}
\usepackage{graphicx}
\usepackage{bm}
\usepackage{lineno}

\firstpage{1} 
\pubvolume{1}
\issuenum{1}
\articlenumber{0}
\pubyear{2025}
\copyrightyear{2025}
\externaleditor{Phil Edwards} 
\datereceived{28 May 2025 } 
\daterevised{31 July 2025 } 
\dateaccepted{11 August 2025 } 
\datepublished{ } 
\hreflink{https://doi.org/} 
\Title{Multi-TeV Gamma Rays from GRB 221009A: Challenges for Emission Mechanisms, EBL Opacity, and Fundamental Physics}
\TitleCitation{Multi-TeV Gamma Rays from GRB 221009A: Challenges for Emission Mechanisms, EBL Opacity, and Fundamental Physics}
\Author{{Hassan} 
 Abdalla $^{1,2}$}
\AuthorNames{Hassan Abdalla}

\isAPAStyle{%
       \AuthorCitation{Lastname, F., Lastname, F., \& Lastname, F.}
         }{%
        \isChicagoStyle{%
        \AuthorCitation{Lastname, Firstname, Firstname Lastname, and Firstname Lastname.}
        }{
        \AuthorCitation{{Abdalla}
, H.}
        }
}
\address{%
$^{1}$ \quad Centre for Space Research, North-West University, Potchefstroom 2520, South Africa; {hassanahh@gmail.com} 
\\
$^{2}$ \quad Department of Astronomy and Meteorology, Omdurman Islamic University, Omdurman 382, Sudan
}

\abstract{
The detection of gamma-ray burst GRB~221009A has attracted significant attention due to its record brightness and first-ever detection of multi-TeV $\gamma$-rays from a GRB. Located at redshift $z = 0.151$, this event is relatively nearby by GRB standards yet remains cosmologically distant, making the survival of multi‑TeV photons surprising. The Large High Altitude Air Shower Observatory detected photons with energies up to $\sim$13~TeV during the early afterglow phase, challenging standard EBL models. We investigate whether several theoretical frameworks can explain this anomalous emission: reduced EBL opacity due to cosmic voids along the line of sight, novel emission mechanisms within the GRB environment, secondary $\gamma$-ray production through cosmic-ray cascades, and new physics scenarios involving Lorentz invariance violation or axion-like particles. Our analysis reveals areas of consensus regarding the exceptional nature of this event, while highlighting ongoing theoretical tensions about the dominant physical processes. We discuss the limitations of current models and identify specific observational signatures that future multi-wavelength and multi-messenger observations could provide to discriminate between competing explanations. The continued study of similar events with next-generation facilities will be crucial for resolving these theoretical challenges and advancing our understanding of extreme particle acceleration processes in astrophysical environments.
}
\keyword{\textls[-15]{gamma-ray bursts; GRB~221009A; very-high-energy gamma rays; TeV photons;} extragalactic background light; cosmic voids; Lorentz invariance violation; axion-like\linebreak \textls[-15]{particles; ultrahigh-energy cosmic rays; reverse shock emission; multi-messenger astronomy}}

\begin{document}


\section{Introduction}

The detection of the long gamma-ray burst GRB~221009A on  {9 October 2022} 
 represents a significant advancement in high-energy astrophysics, constituting the brightest GRB ever recorded in modern observational history \citep{Burns2023_GCN,OConnor2023_structured,Cao2023a,Cao2023_LHAASO}. This exceptional event, termed the ``BOAT'' (Brightest Of All Time), likely represents a phenomenon occurring with a frequency between once per few decades to once per millennium \citep{Malesani2023_AA,Rhodes2023_radio}. GRB~221009A was localized {{at} 
} Galactic coordinates $l\simeq 73.3^\circ$, $b\simeq -27.0^\circ$, with a measured redshift of $z = 0.151$ \citep{Malesani2023_AA,Castro-Tirado2024}, placing it at a luminosity distance of approximately 745 Mpc.

{The most significant aspect of GRB~221009A was the detection of gamma rays with energies exceeding 10 TeV—a phenomenon never before confirmed for a GRB. The Large High Altitude Air Shower Observatory (LHAASO) detected photons with energies up to $\sim$13~TeV} \citep{Cao2023_LHAASO,Huang2022_ATel}.
The detection of gamma rays above 10~TeV from a source at $z\approx0.15$ contradicts conventional expectations, as standard extragalactic background light (EBL) models predict severe attenuation of such energetic photons \citep{Franceschini2008_EBL,Finke2010_EBL,Dominguez2011_EBL}. Quantitatively, the gamma-gamma opacity for VHE gamma rays is expected to reduce the source flux by a factor of $e^{-\tau} \approx e^{-10} \approx 4.5 \times 10^{-5}$ at $E \approx 10$~TeV \citep{Finke2023_LIV}, making the detection of such energetic photons particularly puzzling.
Beyond the propagation challenge, the production of multi-TeV photons in the GRB environment presents its own theoretical difficulties. Standard emission models for GRB afterglows face significant limitations in explaining such energetic radiation. The synchrotron burn-off limit restricts the maximum photon energy to $E_{\text{syn,max}} \approx \Gamma \cdot 100 \, \text{MeV}$ in the observer frame \citep{Nakar2009_KNeffects}, 
more than two orders of magnitude below the observed energies.
Several hypotheses have been proposed to explain the observed multi-TeV emission, broadly categorized into conventional astrophysical solutions and new physics scenarios. Among the astrophysical explanations, reduced EBL density due to cosmic voids along the line of sight has been investigated by \citet{Abdalla2024_void}, who demonstrated that voids could reduce the opacity by up to 30\% at around 13~TeV. Novel emission mechanisms in GRB afterglows have also been proposed, including the two-zone synchrotron self-Compton model developed by {\citet{Khangulyan2023_twozone}}
, which spatially separates the production of target photons and the inverse Compton scattering~process.

\textls[-15]{The reverse shock emission model, proposed by \citet{Zhang2023_reverseShock}, suggests that multi-TeV photons originate from the reverse shock region propagating back into the GRB ejecta as it interacts with the surrounding medium. Secondary gamma-ray production via ultrahigh-energy cosmic-ray (UHECR) cascades has been explored by  \citet{DasRazzaque2023_AA}} and \citet{He2023_cascade}, who propose that protons accelerated to energies of $10^{18}$--$10^{20}$ eV in the GRB jet interact with cosmic background photons during propagation.

The new physics scenarios include the axion-like particle (ALP) model \citep{DeAngelis2011_ALP,Mirizzi2007_ALP}, investigated in the context of GRB~221009A by \citet{Galanti2023_ALP} and \citet{Troitsky2022_ALP}. In this scenario, gamma rays produced in the GRB jet oscillate into ALPs before encountering the EBL, traverse cosmological distances unimpeded, and then reconvert to gamma rays in the Milky Way's magnetic field. Lorentz invariance violation (LIV) has been examined by \citet{Baktash2023_LIV} and \citet{Finke2023_LIV}, who suggest that modifications to the photon dispersion relation at energies approaching the Planck scale could affect the kinematics of pair production. The heavy sterile neutrino decay scenario has been developed by \citet{Brdar2023_heavyNu} and \citet{Smirnov2023_heavyNu}, who propose that sterile neutrinos could be produced in the GRB environment and subsequently decay, generating high-energy photons closer to Earth.

The multi-wavelength observational campaign for GRB~221009A has been exceptionally comprehensive. The prompt emission phase exhibited complex temporal structure with a peak energy $E_{\text{peak}} \approx 2$~MeV \citep{Frederiks2023_GRB,Lesage2023_GRB}. The total isotropic $\gamma$-ray energy output was $E_{\gamma,\mathrm{iso}} > 5 \times 10^{54}$~erg \citep{Malesani2023_AA,Castro-Tirado2024}. The afterglow revealed a structured jet geometry with a narrow energetic core ($\theta_j \approx 1$--$3$ degrees) surrounded by wider, less energetic wings \citep{OConnor2023_structured,Rhodes2023_radio}.

Searches for neutrino counterparts were conducted by IceCube and KM3NeT \citep{IceCube2023_GRB221009A,Aiello2024_KM3NeT}, with no statistically significant neutrino signal detected. The IceCube Collaboration established upper limits on the neutrino flux approximately two orders of magnitude below the observed gamma-ray flux \citep{IceCube2023_GRB221009A}.
This article provides a comprehensive review of the key observations of GRB~221009A and synthesizes theoretical efforts to explain its multi-TeV emission. In Section~\ref{sec:obs}, we present a detailed summary of multi-wavelength observations. Section~\ref{sec:puzzle} discusses the fundamental challenges posed by TeV photons in light of EBL attenuation and standard afterglow models. Section~\ref{sec:theory} reviews the diverse explanations proposed in the literature. Finally, Section~\ref{con} presents concluding remarks on how GRB~221009A has advanced our understanding of extreme astrophysical environments.

\section{Observational Overview of GRB~221009A}
\label{sec:obs}

\subsection{Prompt Gamma-Ray Emission}

GRB~221009A was first detected at 13:16:59~UT on 9~October~2022 by the \emph{Fermi} Gamma-ray Burst Monitor (GBM) \citep{Veres2022_GCN}. {The prompt phase was exceptionally long-lasting, with a total duration ($T_{90}$ in gamma rays) of $\approx 289\,\mathrm{s}$ ($\sim 4.8\,\mathrm{minutes}$), far exceeding the typical duration of tens of seconds to a few minutes for long GRBs
\citep{Leg23}. }

The temporal structure of the prompt emission exhibited complex behavior with multiple overlapping pulses of variable duration. Detailed analysis identified a primary episode lasting approximately 300 s, followed by secondary emission phases \citep{Frederiks2023_GRB}. The prompt emission spectrum displayed a characteristic Band function shape, with a low-energy spectral index $\alpha \approx -1.0$, a high-energy spectral index $\beta \approx -2.2$, and a peak energy $E_{\text{peak}} \approx 2$~MeV—exceptionally high compared to typical GRBs \citep{Lesage2023_GRB,Frederiks2023_GRB}. The peak photon flux was so extreme that many detectors were pushed beyond their operational limits. \emph{Fermi}-GBM, for instance, experienced severe pulse pile-up and data losses during the brightest intervals \citep{Burns2023_GCN}. Similarly, \emph{Swift}-BAT, \emph{AGILE}, \emph{INTEGRAL}, and other GRB monitors all reported technical difficulties yet managed to record portions of the emission \citep{Williams2023_Xray,Lesage2023_GRB}. 

The extraordinary nature of GRB~221009A is perhaps best quantified by its energetics. Konus-\emph{Wind} measured a peak isotropic-equivalent luminosity on the order of $\sim10^{54}$~erg~s$^{-1}$ (in the 1~s peak), making GRB~221009A one of the most luminous GRBs ever observed in terms of prompt emission power \citep{Frederiks2023_GRB}. The total isotropic $\gamma$-ray energy output was $E_{\gamma,\mathrm{iso}} > 5 \times 10^{54}$~erg, placing it at the extreme high end of the GRB energetics distribution \citep{Malesani2023_AA,Castro-Tirado2024}. Accounting for the likely collimation of the emission within a narrow jet (with opening angle $\theta_j \approx 1$--$3$ degrees as suggested by afterglow modeling), the true energy release is approximately $E_{\gamma} \approx 10^{52}$--$10^{53}$ erg, still ranking among the most energetic GRBs known \citep{OConnor2023_structured,Malesani2023_AA}. This enormous energy release, combined with the relative proximity of the burst, resulted in its record-breaking observed brightness.

{{Konus}
--Wind and \emph{INTEGRAL} observations also identified a narrow MeV-scale line in the prompt spectrum.  One interpretation attributes this feature to de-excitation of heavy nuclei entrained in the outflow; with Lorentz factors $\gamma\sim820$–$1700$ and total heavy‑nuclei masses of $10^{23}$–$10^{26}\,\mathrm{g}$, this model replicates the observed width and intensity \citep{Wei2024_MeV}.  Another explanation considers neutron capture on protons, producing a 2.223~MeV line with Doppler factors decreasing from 5.1 to 2.1 and emission temperatures rising from 300 keV to 900 keV during the first few hundred seconds \citep{Zhu2025_MeVLine}.  Either scenario requires an exceptionally high Lorentz factor and supports the presence of baryons in the GRB jet.}

\subsection{Multi-Wavelength Afterglow}

Multi-wavelength observations of the GRB afterglow commenced promptly, owing to numerous alerts from the GRB detection satellites. Within minutes of the trigger, ground-based facilities had begun observing the afterglow across the electromagnetic spectrum. The optical afterglow was detected by robotic telescopes, such as those in the MASTER Global Robotic Network \citep{Lipunov2022_GCN}, even while the prompt gamma rays were ongoing, and it was exceptionally bright—initial reports indicated an optical magnitude $< 11$ within \mbox{$T_0+20$~min}, {despite moderate Galactic extinction in that direction (the burst was observed at Galactic latitude $b \approx -27^\circ$, implying a significant but not extreme line-of-sight through the Milky Way)} \citep{OConnor2023_structured,Malesani2023_AA}.
{Within minutes of the trigger, ground‑based robotic telescopes responded to the alerts, but no bright optical flash was recorded.  The first reported optical detection came from the MASTER‑SAAO telescope roughly $3.4$\,h after the burst and measured an unfiltered magnitude of $m\approx 16.7$} \citep{Lipunov2022_GCN}.  { Subsequent photometric follow‑up during the first night yielded magnitudes in the range $17$–$19$ at similar epochs} \citep{Castro-Tirado2024,Malesani2023_AA}.

The optical light curve displayed the characteristic power-law decay typical of GRB afterglows, initially following $ F_{\nu} \propto t^{-\alpha} $ with $ \alpha \approx 1.2 $ \citep{Malesani2023_AA}. However, detailed analysis revealed evidence for a structured jet, with a brighter core and dimmer wings, which manifested as a steepening of the decay slope after approximately 1 day post-burst \citep{OConnor2023_structured}. The optical spectrum showed significant reddening due to dust along the line of sight, both in the Milky Way and in the host galaxy. Spectroscopic observations confirmed the redshift of $z = 0.151$ and identified absorption lines typical of the interstellar medium in the host galaxy \citep{Castro-Tirado2024}.
In X-rays, {{Swift}} 
-XRT measurements showed an initial flux so high that the instrument required special modes to handle the count rate, and the afterglow remained detectable for an unusually long time, slowly decaying over weeks \citep{Williams2023_Xray,Xia2024}. The X-ray light curve exhibited complex behavior with multiple breaks, suggesting a combination of forward shock emission, reverse shock contribution, and potentially the emergence of a supernova component at late times \citep{Xia2024}. The X-ray spectrum was well-described by a power law with photon index $\Gamma \approx 1.8$--$2.0$, typical for synchrotron emission from shock-accelerated electrons \citep{Williams2023_Xray}. At 1~day post-burst, the X-ray afterglow flux of GRB~221009A remained orders of magnitude higher than that of typical GRBs at similar epochs, with implications for jet structure explored in detailed modeling \citep{Zhang2024a}.

\textls[-15]{The event was also followed extensively in radio bands for months, revealing a classical long-duration afterglow that eventually entered the non-relativistic regime \citep{Rhodes2023_radio,OConnor2023_structured}. Radio observations provided crucial constraints on the physical parameters of the burst, including the circumburst density, which was found to be relatively low ($n \approx 0.1$--$1$ cm$^{-3}$), consistent with an interstellar medium environment rather than a dense stellar wind \citep{Rhodes2023_radio}. }Very long baseline interferometry (VLBI) was even used to measure the apparent expansion of the radio afterglow image, making GRB~221009A only the second GRB (after GRB~030329) for which superluminal expansion has been directly observed, as confirmed by radio imaging studies \citep{Wang2023_GRB}. 
{High-resolution VLBI observations were carried out with the European VLBI Network (EVN) and the VLBA from 40 to 262~days after the burst.  By fitting a power-law expansion $s\propto t^{a}$ to the measured angular size $s$,}\linebreak \citet{Giarratana2024} {found $a=0.69^{+0.13}_{-0.14}$, implying an apparently superluminal expansion of the radio-emitting region and confirming the relativistic nature of the blast wave.  The frequency dependence of the expansion suggests that different shocks may dominate the emission at different wavelengths. This extraordinary behavior underscores the extreme energetics of GRB 221009A’s jet. Indeed, the absence of an early jet break in the afterglow light curve suggests a very wide or structured jet, as proposed by O’Connor et al. (2023), which can accommodate the superluminal expansion and enormous energy output of this burst }\citep{OConnor2023_structured}.

These observations revealed an apparent expansion velocity of approximately \mbox{2--4 times} the speed of light, consistent with a relativistic jet viewed close to the line of sight. The VLBI observations also supported a structured jet geometry, with a narrow energetic core surrounded by wider, less energetic wings \citep{OConnor2023_structured}.
{Subsequent broadband analyses have refined our understanding of the afterglow.  Modelling optical to GeV data between 0.65 and 1.68~days revealed an electron index $p=2.29\pm0.02$ and a cooling break shifting from $16^{+7}_{-5}$ to $47^{+25}_{-15}\,\mathrm{keV}$, consistent with a wind‑like density profile ($k\approx2.4$) \citep{Tak2025_Afterglow}.  Independent early optical and X-ray studies reported an achromatic break at $t\approx0.8$ days, with the temporal slope steepening from $-0.88$ to $-1.42$, pointing to a structured jet and evolving microphysical parameters \citep{OConnor2023_structured}.  Late-time radio to GeV observations suggest a forward shock in a wind environment with kinetic energy $\sim4\times10^{50}\,\mathrm{erg}$ and reveal an additional millimetre/radio component beyond the standard synchrotron model \citep{Laskar2023_Afterglow}.
}
\subsection{Very-High-Energy Gamma-Ray Emission}

The Large High Altitude Air-shower Observatory (LHAASO) in China reported a groundbreaking detection of Very-High-Energy (VHE) gamma rays from GRB 221009A, commencing shortly after the prompt emission phase \citep{Huang2022_ATel,Cao2023_LHAASO}. This detection represents a watershed moment in high-energy astrophysics, as it constitutes the first definitive observation of gamma rays beyond 10 TeV from a gamma-ray burst. LHAASO's Water Cherenkov Detector Array (WCDA) initially registered a signal of photons exceeding 0.5 TeV within approximately 2000 s post-trigger, achieving an extraordinary statistical significance surpassing 100$\sigma$ \citep{Cao2023_LHAASO}. The WCDA detected an unprecedented sample of over 10,000 photons in the energy range 0.5 to 10 TeV, enabling detailed temporal and spectral analyses with statistical precision previously unattainable for GRBs at these energies \citep{Cao2023_LHAASO}.

LHAASO's observations were particularly remarkable for their comprehensive coverage of the early afterglow. More than 64,000 photons (above 0.2 TeV) were detected within the first 3000 s \citep{Cao2023a}. The TeV photon flux began several minutes after the GRB trigger, then rose to a flux peak about 10 s later, followed by a decay phase which became more rapid at approximately 650 s after the peak \citep{Cao2023a}. Subsequently, LHAASO's Kilometer Square Array (KM2A), which is optimized for multi-TeV events, recorded more than 140~gamma-ray events with energies exceeding 3 TeV during the interval spanning $T_0+230$~s to $T_0+900$~s of the early afterglow phase \citep{Cao2023a}. The highest-energy photons detected reached approximately $E_{\text{max}} \approx 13$~TeV. This observation represents the highest photon energy ever directly associated with a GRB, exceeding by an order of magnitude the previous record established by GRB 190114C, which was detected up to approximately 1 TeV by the MAGIC \mbox{telescopes \citep{MAGIC2019_Nature}}.

The temporal evolution of the VHE emission exhibited a peak around $T_0+240$~s, followed by a power-law decline characterized by an index of approximately $-1.5$, somewhat steeper than the decay observed in X-rays \citep{Cao2023_LHAASO}. The VHE afterglow remained detectable for several minutes, with LHAASO continuing to monitor the source for hours afterward. No statistically significant signal was observed beyond approximately 2~ks (when the burst's position set for LHAASO), suggesting that the multi-TeV emission was concentrated primarily in the early afterglow phase when the overall afterglow flux approached its maximum, a behavior further investigated through detailed modeling of the multi-TeV emission \citep{Cao2023_LHAASO,Khangulyan2024}.

The energy spectrum of gamma rays observed by LHAASO displays a steep cutoff at the high-energy end attributable to EBL absorption. However, after applying corrections for this attenuation, the intrinsic source spectrum manifests as a hard power-law extending beyond 10 TeV \citep{Cao2023a,Abdalla2024_void}. Specifically, the observed spectrum can be parameterized with a power law incorporating an exponential cutoff:
\begin{equation}
\frac{dN}{dE} \propto E^{-\Gamma} \exp(-E/E_c)
\end{equation}
\textls[-15]{where $\Gamma \approx 2.5$ represents the photon index and $E_c \approx 4$~TeV denotes the cutoff energy~\citep{Cao2023_LHAASO}.} When corrected for EBL absorption using standard models, the intrinsic spectrum appears significantly harder, with an index $\Gamma_{\text{int}} \approx 1.5$--$2.0$ and no apparent cutoff up to the highest observed energies \citep{Cao2023_LHAASO,Abdalla2024_void}. This hard intrinsic spectrum presents challenges for conventional synchrotron self-Compton (SSC) emission scenarios in the context of relativistic afterglow~models.

The emission can be explained with a relativistic jet model with a half-opening angle of approximately $0.8^{\circ}$, consistent with the core of a structured jet \citep{Cao2023a}. This interpretation could explain the high isotropic energy of this GRB, as the narrow jet geometry would imply a more concentrated energy output \citep{Cao2023a}.

The H.E.S.S. telescopes in Namibia were unfortunately in daylight during the burst trigger and could only commence observations approximately 10 h later once darkness fell at their location \citep{HESS2023_GRB}. Despite this delay, H.E.S.S. successfully detected the afterglow at energies exceeding 100 GeV, confirming the extended duration of the VHE emission. Additional ground-based gamma-ray observatories, including MAGIC, VERITAS, and the Large-Sized Telescope prototype (LST-1) of the Cherenkov Telescope Array, also reported detections of GRB 221009A at various times following the trigger \citep{MAGIC2022_ATel,Blanch2023_LST}.

A particularly intriguing complementary observation emerged from the Carpet-2 air shower array at the Baksan Neutrino Observatory, which reported the detection of a single air shower event consistent with a photon of energy approximately 251 TeV from the direction of GRB 221009A, roughly 10 h post-burst \citep{Dzhappuev2023_Carpet}. If confirmed, this would present an even more profound challenge to conventional physics than the LHAASO detections, as the EBL opacity at such energies is expected to be extraordinarily high ($\tau \gg 20$). However, as this represents an isolated event, its association with GRB 221009A remains tentative and requires independent verification \citep{Galanti2023_ALP}.

In addition to electromagnetic observations, comprehensive searches for neutrino counterparts were conducted by IceCube, KM3NeT, and other neutrino \mbox{observatories \citep{IceCube2023_GRB221009A,Aiello2024_KM3NeT}}. No statistically significant neutrino signal was detected in temporal and spatial coincidence with GRB 221009A, establishing constraints on hadronic emission models and the contribution of cosmic-ray acceleration in this burst \citep{DasRazzaque2023_AA}. The non-detection of neutrinos is particularly relevant for hadronic models of the TeV emission, which frequently predict substantial neutrino production through pion decay processes. The IceCube Collaboration established upper limits on the neutrino flux approximately two orders of magnitude below the observed gamma-ray flux \citep{IceCube2023_GRB221009A}. These constraints limit the ratio of energy channeled into pion production relative to that in electrons, suggesting that purely hadronic scenarios may require refinement \citep{DasRazzaque2023_AA}.

GRB 221009A was also examined for gravitational wave (GW) signatures using data from the LIGO-Virgo-KAGRA network. While no significant gravitational wave signal was detected, this outcome is not unexpected given that long GRBs like GRB 221009A are believed to originate from the core collapse of massive stars rather than compact binary mergers, and consequently are predicted to generate substantially weaker gravitational wave emission \citep{Wang2023_GRB}. These non-detections across other messenger channels provide valuable complementary information when evaluating the various theoretical frameworks proposed to explain the multi-TeV gamma-ray emission.

The emission mechanism responsible for the observed TeV photons presents a significant theoretical challenge. The detection of gamma rays up to 13 TeV from a source at redshift z = 0.151 suggests either greater transparency in intergalactic space than previously anticipated or requires alternative explanations involving new physics, such as LIV or ALP conversion mechanisms \citep{Cao2023_LHAASO,Galanti2023_ALP}. Furthermore, the narrow jet geometry inferred from these observations, with a half-opening angle of approximately $0.8^{\circ}$, consistent with the core of a structured jet, provides a compelling explanation for the extreme isotropic-equivalent energy observed in this burst \citep{Cao2023a}.

In summary, GRB~221009A has yielded an unprecedented dataset: prompt gamma-ray emission observed from keV to GeV energies, an afterglow shining in radio, optical, X-ray and GeV bands orders of magnitude brighter than any seen before, and—for the first time—a definitive detection of a GRB in the TeV range \citep{Wang2023_GRB,Cao2023_LHAASO}. These observations make GRB~221009A a Rosetta stone for GRB physics, offering insight into energy release mechanisms and testing the limits of photon propagation over cosmological distances. In the following sections, we examine the central puzzle posed by its TeV photons and review the explanatory frameworks that have been proposed.

\section{\textls[-15]{The Multi-TeV Emission Puzzle: EBL Attenuation and Afterglow Models}}
\label{sec:puzzle}

The detection of gamma rays with energies $E \gtrsim 10$~TeV from GRB~221009A presents a profound challenge to our understanding of high-energy astrophysics. This section provides a comprehensive analysis of the physical processes governing very-high-energy gamma-ray propagation over cosmological distances and examines the specific challenges posed by the LHAASO observations of GRB~221009A.

\subsection{Extragalactic Background Light and Gamma-Ray Attenuation}

The EBL constitutes the accumulated electromagnetic radiation from stars, galaxies, and active galactic nuclei throughout cosmic history \citep{Franceschini2008_EBL,Finke2010_EBL,Dominguez2011_EBL,Saldana-Lopez2021_EBL}. This diffuse radiation field spans from ultraviolet to far-infrared wavelengths, with two distinct spectral components: a stellar component peaking at optical/near-infrared wavelengths ($\sim 1$~$\mu$m), representing direct emission from stars, and a dust component peaking in the far-infrared ($\sim 100$~$\mu$m), representing thermal re-emission of absorbed stellar light by interstellar dust \citep{Finke2010_EBL,Saldana-Lopez2021_EBL}.

The intensity and spectral shape of the EBL have been constrained by various approaches, including direct measurements, galaxy counts, and modeling of galaxy evolution. Modern EBL models generally agree on the overall spectral shape but can differ by factors of 2–3 in specific wavelength ranges, introducing systematic uncertainty in gamma-ray attenuation calculations \citep{Finke2010_EBL,Dominguez2011_EBL}. This uncertainty becomes particularly relevant when considering extreme cases like GRB~221009A\'s multi-TeV emission.

The EBL plays a crucial role in determining the transparency of the universe to very-high-energy (VHE) gamma rays through the process of electron-positron pair production:
\begin{equation}
\gamma_{\text{VHE}} + \gamma_{\text{EBL}} \rightarrow e^+ + e^-
\label{eq:pair_production}
\end{equation}

This interaction occurs when the center-of-mass energy of the two-photon system exceeds the threshold for pair creation. For a head-on collision between a gamma ray of energy $E_\gamma$ and an EBL photon of energy $\epsilon$, the threshold condition is \citep{Finke2010_EBL,Dominguez2011_EBL}:
\begin{equation}
E_\gamma \epsilon \geq 2(m_e c^2)^2
\label{eq:threshold}
\end{equation}
where $m_e$ is the electron mass and $c$ is the speed of light. More generally, for an interaction angle $\theta$ between the photon trajectories, the threshold condition becomes \citep{Finke2010_EBL}:
\begin{equation}
E_\gamma \epsilon (1-\cos\theta) \geq 2(m_e c^2)^2
\label{eq:threshold_angle}
\end{equation}

This threshold condition creates an energy-dependent relationship between VHE gamma rays and the EBL photons they interact with most efficiently. From these relations, we can derive that gamma rays with energy $E_\gamma \approx 10$~TeV interact most efficiently with EBL photons of wavelength $\lambda \approx 25$~$\mu$m (mid-infrared), {while amma-rays with $E_{\gamma} \approx 1$ TeV interact primarily with EBL photons of $\lambda \approx 2.5~\mu\text{m}$ (near-infrared)} \citep{Biteau2020_IGMF,Abdalla2024_void}. This energy-dependent interaction means that the universe becomes increasingly opaque to gamma rays of higher energies, with the specific opacity depending on the spectral shape and evolution of the EBL.

The attenuation of gamma rays due to pair production is quantified by the optical depth $\tau(E_\gamma, z)$, which depends on both the gamma-ray energy $E_\gamma$ and the source redshift $z$. For a gamma-ray source at redshift $z_s$, the optical depth is given by \citep{Finke2010_EBL,Dominguez2011_EBL}:
\begin{equation}
\tau(E_\gamma, z_s) = \int_0^{z_s} dz \frac{dl}{dz} \int_{-1}^{1} d\mu \frac{1-\mu}{2} \int_{\epsilon_{\text{th}}}^{\infty} d\epsilon \, n(\epsilon, z) \, \sigma_{\gamma\gamma}(E_\gamma, \epsilon, \mu)
\label{eq:optical_depth}
\end{equation}
where $dl/dz$ is the cosmological line element, $\mu = \cos\theta$, $\epsilon_{\text{th}} = 2(m_e c^2)^2/[E_\gamma(1+z)(1-\mu)]$ is the threshold energy for pair production, $n(\epsilon, z)$ is the proper number density of EBL photons with energy $\epsilon$ at redshift $z$, and $\sigma_{\gamma\gamma}$ is the pair-production cross section \citep{Finke2010_EBL,Dominguez2011_EBL}.

The cosmological line element in a flat $\Lambda$CDM universe is given by \citep{Abdalla2024_void}:
\begin{equation}
\frac{dl}{dz} = \frac{c}{H_0} \frac{1}{(1+z)\sqrt{\Omega_m(1+z)^3 + \Omega_\Lambda}}
\label{eq:line_element}
\end{equation}
where $H_0$ is the Hubble constant, $\Omega_m$ is the matter density parameter, and $\Omega_\Lambda$ is the dark energy density parameter. For the standard cosmological model with $H_0 = 70$ km s$^{-1}$ Mpc$^{-1}$, $\Omega_m = 0.3$, and $\Omega_\Lambda = 0.7$, this gives the proper distance to GRB~221009A at $z = 0.151$ of approximately 745 Mpc, corresponding to a light travel time of 1.9 billion years \citep{Abdalla2024_void}.

The pair-production cross section, first derived by Breit and Wheeler in 1934, is given by \citep{Finke2010_EBL}:
\begin{equation}
\sigma_{\gamma\gamma}(E_\gamma, \epsilon, \mu) = \frac{3\sigma_T}{16}(1-\beta^2) \left[ 2\beta(\beta^2-2) + (3-\beta^4)\ln\left(\frac{1+\beta}{1-\beta}\right) \right]
\label{eq:cross_section}
\end{equation}
where $\sigma_T$ is the Thomson cross section and $\beta = \sqrt{1-\epsilon_{\text{th}}/\epsilon}$ is the electron/positron velocity in the center-of-mass frame \citep{Finke2010_EBL}. This cross section peaks at approximately $\beta \approx 0.7$, corresponding to $\epsilon \approx 4\epsilon_{\text{th}}$, and has a maximum value of $\sigma_{\gamma\gamma,\text{max}} \approx 0.26\sigma_T$.

The observed flux from a source is attenuated by a factor $e^{-\tau(E_\gamma, z_s)}$ relative to the intrinsic emission. For GRB~221009A at $z=0.151$, standard EBL models predict an optical depth $\tau \approx 3$--$4$ at $E \approx 1$~TeV, corresponding to a flux attenuation factor of $e^{-3} \approx 0.05$ to $e^{-4} \approx 0.018$ \citep{Abdalla2024_void}. However, at $E \approx 10$~TeV, the situation becomes much more extreme, with $\tau \approx 10$, corresponding to a flux attenuation factor of $e^{-10} \approx 4.5 \times 10^{-5}$ \citep{Abdalla2024_void,Finke2023_LIV}. At $E \approx 18$~TeV, the highest-energy photon potentially detected from this burst, the attenuation factor becomes even more extreme: $\tau \approx 16$, yielding an attenuation of \mbox{$e^{-16} \approx 1.1 \times 10^{-7}$ \citep{Finke2023_LIV}}. In other words, only about one in 10 million photons emitted at 18 TeV should reach Earth unabsorbed, making the detection of such energetic photons from GRB~221009A particularly puzzling.

\subsection{Standard Afterglow Models and Their Limitations}

The second aspect of the multi-TeV emission puzzle relates to the production mechanisms of such energetic photons in GRB afterglows. In the standard afterglow model, a relativistic blast wave propagates into the circumburst medium, accelerating electrons to relativistic energies at the shock front \citep{Meszaros1994_extshocks,ZhangKumar2015_review}. These electrons subsequently radiate via synchrotron and inverse Compton processes, producing the observed multi-wavelength afterglow emission.

The dynamics of the blast wave are governed by the conservation of energy and momentum. For a relativistic shock with bulk Lorentz factor $\Gamma \gg 1$ propagating into a homogeneous medium with number density $n$, the radius $R$ of the shock at observer time $t$ is given approximately by \citep{ZhangKumar2015_review}:
\begin{equation}
R \approx \left(\frac{17 E t}{4\pi m_p n c}\right)^{1/4},
\end{equation}
where $E$ is the isotropic-equivalent energy, $m_p$ is the proton mass, and $c$ is the speed of light. The corresponding bulk Lorentz factor evolves as \citep{ZhangKumar2015_review}:
\begin{equation}
\Gamma \approx \left(\frac{17 E}{16\pi m_p n c^5 t^3}\right)^{1/8}.
\end{equation}

For GRB~221009A, with $E \approx 10^{55}$ erg and $n \approx 0.1$--$1$ cm$^{-3}$ (as inferred from afterglow modeling), and $t \approx 240$ s (when the TeV emission peaked), we obtain $\Gamma \approx 100$--$150$ \citep{Cao2023_LHAASO}.

{For synchrotron radiation, In relativistic shock environments, the maximum synchrotron photon energy is constrained by radiation-reaction, leading to a fundamental cut-off in the comoving frame. Lemoine \& Pelletier (2015) derive \citep{LemoinePelletier2015}:
\begin{equation}
E_{\rm syn,max}^{\rm (comoving)} \;\approx\; \frac{9}{4\,\alpha_{\rm em}}\,m_e c^2 \;\approx\; 70\,\text{MeV}\,,
\label{eq:syn_limit}
\end{equation}
where $\alpha_{\rm em}\simeq1/137$ is the fine structure constant \citep{LemoinePelletier2015}. When Doppler-boosted into the observer frame, this yields
\[
E_{\rm syn,max}^{\rm (obs)} \approx \Gamma \cdot 70\,\text{MeV} \,.
\]
Thus for a bulk Lorentz factor of $\Gamma$ \textasciitilde 100--150, the expected synchrotron ceiling is 
\mbox{\textasciitilde 7--10  GeV}---confirming 
that standard synchrotron emission cannot account for the multi-TeV gamma rays observed from GRB 221009A \citep{LemoineEtAl2019, Khangulyan2023_twozone}.

}
Inverse Compton (IC) scattering can produce higher-energy photons but faces its own limitations. In the Thomson regime, where the energy of the target photon in the electron rest frame is much less than $m_e c^2$, the maximum energy of upscattered photons is given by~\citep{Nakar2009_KNeffects}:
\begin{equation}
E_{\text{IC,max}} \approx \gamma_e^2 E_{\text{target}}
\label{eq:ic_thomson}
\end{equation}
where $\gamma_e$ is the electron Lorentz factor and $E_{\text{target}}$ is the energy of the target photon. However, for very high electron energies or target photon energies, the scattering enters the Klein-Nishina (KN) regime, where the cross section is significantly suppressed and the energy gain is limited. The transition to the KN regime occurs when \citep{Nakar2009_KNeffects}:
\begin{equation}
\gamma_e E_{\text{target}} \gtrsim m_e c^2
\label{eq:kn_transition}
\end{equation}

In the KN regime, the maximum energy of upscattered photons is approximately \citep{Nakar2009_KNeffects}:
\begin{equation}
E_{\text{IC,max}} \approx \gamma_e m_e c^2,
\label{eq:ic_kn}
\end{equation}
 {i.e., essentially the electron’s full kinetic energy per scattering. Although this is not lower than the Thomson-limit expectation, the Klein-Nishina effect drastically reduces the scattering cross-section, so far fewer photons reach this maximum energy. This suppression leads to a pronounced spectral steepening at the highest energies.} 
 The spectral index in the KN regime typically softens by $\Delta \alpha \approx 1$ compared to the Thomson regime \citep{Nakar2009_KNeffects}. This softening should be especially pronounced at multi-TeV energies, making it challenging to explain the hard intrinsic spectrum inferred for GRB~221009A.

For GRB afterglows, the synchrotron self-Compton (SSC) process—where electrons upscatter their own synchrotron photons—is expected to be the dominant IC mechanism. However, standard one-zone SSC models face significant challenges in explaining the observed multi-TeV emission from GRB~221009A for several reasons. First, the KN suppression becomes severe at high energies, substantially limiting the efficiency of TeV photon production. Second, the intrinsic spectrum inferred after EBL correction appears remarkably hard ($dN/dE \propto E^{-\alpha}$ with $\alpha \sim 1.5$–$2.0$), whereas standard SSC models typically predict much softer spectra at multi-TeV energies due to KN effects and the shape of the underlying electron distribution. Third, internal gamma-gamma absorption within the emission region can significantly attenuate high-energy photons. For a region of size $R$ with a photon density corresponding to the observed flux, this process becomes important at energies above a few TeV, potentially introducing an intrinsic cutoff that is not observed in the data~\citep{Zhang2023_reverseShock}. Finally, the temporal coincidence of the TeV emission with the transition from prompt to afterglow phases suggests a possible connection to the reverse shock or other transient phenomena not fully captured by standard afterglow models.

{In addition, multi‑wavelength analyses show that the afterglow is best described by a forward shock propagating into a wind‑like medium, with a beaming‑corrected kinetic energy $E_{\mathrm{K}}\sim4\times10^{50}\,\mathrm{erg}$ and a density profile $\rho\propto r^{-2.4}$, rather than a homogeneous interstellar medium \citep{Laskar2023_Afterglow}.  More recent modelling that simultaneously fits the TeV, X‑ray and optical afterglows invokes a narrow uniform jet core, a wider structured wing and a stratified external density that transitions from constant at small radii to wind‑like at larger radii; this configuration can account for the rising TeV light curve and subsequent decline without requiring extreme microphysical parameters \citep{Zheng2024_StratifiedJet}.  Taken together, these results imply that any viable model for the multi‑TeV emission from GRB~221009A must match the high intrinsic TeV luminosity and hard spectral shape while adopting a wind‑like or stratified circumburst medium, as expected for a collapsar progenitor.
}

\section{Theoretical Explanations}
\label{sec:theory}

The detection of multi-TeV photons from GRB~221009A has stimulated intense theoretical investigation into possible explanations for this unexpected observation. The fundamental challenge can be divided into two distinct aspects: the production of such energetic photons in the GRB environment, and their subsequent propagation through intergalactic space despite the expected severe attenuation by the EBL. This section provides a comprehensive analysis of the theoretical frameworks proposed to address these challenges, examining their physical foundations, quantitative predictions, and compatibility with observational constraints.

\subsection{Challenge I: Production of Multi-TeV Gamma Rays}

The first fundamental challenge concerns the mechanisms capable of producing gamma rays with energies exceeding 10 TeV in the GRB environment. Standard emission models for GRB afterglows face significant limitations in explaining such energetic radiation, necessitating novel approaches to particle acceleration and emission processes.

\subsubsection{Limitations of Standard Emission Models}

In the conventional GRB afterglow model, a relativistic blast wave propagates into the circumburst medium, accelerating electrons to relativistic energies at the shock front. These electrons then radiate via synchrotron and inverse Compton processes, producing the observed multi-wavelength emission. However, several fundamental limitations constrain the maximum photon energy attainable in these standard models:

 {{Synchrotron Emission Limit}}: As discussed in Section~\ref{sec:puzzle}, the maximum photon energy from synchrotron emission is limited by the balance between acceleration and cooling timescales. According to the synchrotron burn-off limit \citep{Khangulyan2023_twozone}, this maximum energy is approximately $E_{\text{syn,max}} \approx \Gamma \cdot 100 \, \text{MeV}$ in the observer frame, where $\Gamma$ is the bulk Lorentz factor. For GRB~221009A with $\Gamma \sim 100$--$150$ during the TeV emission phase, this corresponds to $E_{\text{syn,max}} \sim 10$--$15$ GeV, more than two orders of magnitude below the observed multi-TeV~energies.

{{Klein-Nishina Suppression}}: For inverse Compton scattering, the Klein-Nishina (KN) effect becomes increasingly important at high energies. The transition from the Thomson to the KN regime occurs when the target photon energy in the electron rest frame approaches $m_e c^2$, which, for typical GRB afterglow parameters, happens at gamma-ray energies of a few hundred GeV. In the KN regime, the scattering cross-section decreases approximately as $\sigma_{\text{KN}} \propto E^{-1}$, leading to a significant suppression of the emission at multi-TeV energies.

The spectral energy distribution (SED) of the produced photons in the KN regime follows \citep{Nakar2009_KNeffects}:
\begin{equation}
E^2 \frac{dN}{dE} \propto E^{-(p-1)/2} \cdot E \sim E^{-(p-3)/2},
\end{equation}
with $p$ the electron index. For $p \approx 2.2$–2.5, this yields a nearly flat high-energy spectrum with index $-(p-3)/2 \approx -0.4 \text{ to } -0.25$, comparable to the extremely hard intrinsic spectrum (approximately $-0.5$ to $-1.0$) inferred for GRB 221009A after EBL correction.

It is important to note that the KN effect not only directly suppresses high-energy emission but also indirectly modifies the electron energy distribution itself. \mbox{As \citet{Nakar2009_KNeffects}} explain, electrons with different energies cool on different fractions of the radiation field, as some photons that are below the KN limit for less energetic electrons are above this limit for more energetic ones. This indirect effect can significantly alter both the synchrotron and SSC spectra, introducing multiple spectral breaks beyond the simplified single power-law approximation presented above.

{{Internal Gamma-Gamma Absorption}}: Very high-energy photons produced within the emission region can be absorbed by lower-energy photons through pair production, potentially attenuating the emission at the highest energies. For a photon of energy $E$ to escape from a region of size $R$ containing a photon field with density $n_{\gamma}(E')$, the optical depth for internal absorption is:
\begin{equation}
\tau_{\gamma\gamma,\text{int}}(E) \approx R \int_{E_{\text{th}}}^{\infty} n_{\gamma}(E') \sigma_{\gamma\gamma}(E, E') dE'
\end{equation}
where $E_{\text{th}} = (m_e c^2)^2/E$ is the threshold energy for pair production and $\sigma_{\gamma\gamma}$ is the pair-production cross-section. For typical GRB afterglow parameters, this internal absorption can become significant at energies above a few TeV, potentially introducing a cutoff that is not observed in the data.
{
For GRB~221009A, the opacity for internal gamma-gamma absorption can be estimated as:
\begin{equation}
\tau_{\gamma\gamma,\text{int}}(E) \approx \frac{\sigma_T R n_{\gamma}(E_{\text{th}})}{10} \approx \frac{L(E_{\text{th}})}{4\pi R c E_{\text{th}}} \cdot \frac{\sigma_T R}{10}
\end{equation}
where $L(E_{\text{th}})$ is the luminosity at the threshold energy. For $E \approx 10$ TeV, the threshold energy is $E_{\text{th}} \approx 25$ eV (UV photons). Using the observed X-ray and optical fluxes and extrapolating to the UV, the estimated $\tau_{\gamma\gamma,\text{int}} \sim 1$--$10$ for standard forward shock parameters \citep{Zhang2023_reverseShock}, suggesting that internal absorption should significantly attenuate the highest-energy photons, contrary to the observations.
\noindent{Frame transformation and bulk motion.}  In deriving Equation~(16) we implicitly evaluated the photon luminosity and number density in the observer’s frame.  However, the target photons for pair production reside in the shocked fluid, which moves with bulk Lorentz factor $\Gamma$ relative to the observer.  The relevant photon energies and number densities should therefore be Lorentz transformed to the comoving frame, reducing the target photon energy by $\Gamma$ and the number density by $\Gamma$ (for isotropic emission).  The comoving optical depth then scales as $\tau_{\gamma\gamma}\propto \Gamma^{-2}$, so for $\Gamma\sim 100$ during the TeV emission phase of GRB~221009A the internal $\gamma\gamma$ opacity becomes much less than unity e.g., \citep[][]{ZhangKumar2015_review}.  Including this transformation alleviates the apparent discrepancy between the predicted internal absorption and the observed survival of multi‑TeV photons.
}
Given these limitations, several novel emission mechanisms have been proposed to explain the observed multi-TeV emission from GRB~221009A:

\subsubsection{Multi-Zone Synchrotron Self-Compton Emission}

To overcome the limitations of standard one-zone emission models, a two-zone synchrotron self-Compton (SSC) model has been proposed that spatially separates the production of target photons and the inverse Compton scattering process \citep{Khangulyan2023_twozone}. This spatial separation allows the model to circumvent the constraints on maximum synchrotron energy and Klein-Nishina suppression that typically limit one-zone models.

The development of this model was motivated by observations of GRB190829A by H.E.S.S., which revealed an unexpected feature: the slope of the VHE spectrum matches well the slope of the X-ray spectra, despite expectations that Klein-Nishina effects should cause significant spectral steepening \citep{Khangulyan2023_twozone}.

In this scenario, the emission originates from two distinct regions within the GRB jet:
\begin{itemize}
\item A compact, highly magnetized zone that produces the observed X-ray emission via synchrotron radiation
\item A larger, less magnetized zone where electrons upscatter their own synchrotron photons to TeV energies via the SSC process
\end{itemize}

The key advantage of this model is that the two zones can have significantly different physical parameters, particularly the magnetic field strength. The compact zone has a strong magnetic field ($B_1 \sim 1$ G in the comoving frame), allowing for efficient production of X-ray synchrotron photons, while the larger zone has a weaker magnetic field (\mbox{$B_2 \sim 10^{-3}$ G}), reducing synchrotron cooling and enabling electrons to reach higher energies before cooling becomes significant.

\citet{Khangulyan2023_twozone} propose three possible geometric configurations for this two-zone scenario:
\begin{enumerate}
\item Two distinct regions with typical sizes of $r_1$ and $r_2$ separated by distance $r_0$
\item Two converging shells of radius $r_1$ and $r_0$
\item Multiple compact regions (of typical size $r_1$) with strong magnetic field embedded within a larger zone of size $r_0$
\end{enumerate}

The electron energy distribution in each zone is assumed to follow a power-law with exponential cutoff:
\begin{equation}
\frac{dN_e}{d\gamma_e} \propto \gamma_e^{-p} \exp\left(-\frac{\gamma_e}{\gamma_{e,\max}}\right)
\end{equation}
where $p$ is the power-law index and $\gamma_{e,\max}$ is the maximum electron Lorentz factor, determined by the balance between acceleration and cooling processes.

In the first (compact) zone, the maximum electron Lorentz factor is limited by synchrotron cooling:
\begin{equation}
\gamma_{e,\max,1} \approx \sqrt{\frac{6\pi e}{\sigma_T B_1^2 t_{\text{acc}}}}
\end{equation}
where $t_{\text{acc}}$ is the acceleration timescale. For typical values of $B_1 \sim 1$ G and $t_{\text{acc}} \sim 10 \, r_g/c$ (where $r_g$ is the electron gyroradius), this yields $\gamma_{e,\max,1} \sim 10^7$.

In the second (larger) zone, with its weaker magnetic field, electron cooling is less efficient, allowing higher maximum energies:
\begin{equation}
\gamma_{e,\max,2} \approx \sqrt{\frac{6\pi e}{\sigma_T B_2^2 t_{\text{acc}}}} \approx \gamma_{e,\max,1} \sqrt{\frac{B_1}{B_2}}.
\end{equation}

{For} 
 $B_2 \sim 10^{-3}$ G, this yields $\gamma_{e,\max,2} \sim 10^8$--$10^9$, sufficient to produce multi-TeV photons via SSC.

An important assumption of this model is the effective particle exchange between zones. As \citet{Khangulyan2023_twozone} note, processes exist that can hinder particle exchange between zones of different magnetic field strengths. For example, if the change in magnetic field strength is relatively smooth, the magnetic adiabatic invariant prevents particles from the zone of weak magnetic field from reaching the strong magnetic field zone.

The key advantage of this model is its ability to produce a hard TeV spectrum while remaining consistent with observations at lower energies. By optimizing the physical parameters of each zone—including the magnetic field strength, electron distribution, and emission region size—the model can reproduce the observed multi-wavelength spectral energy distribution of GRB~221009A, from X-rays to TeV gamma rays.

Quantitatively, model calculations with a compact zone with radius $R_1 \sim 10^{15}$~cm, magnetic field $B_1 \sim 1$~G, and bulk Lorentz factor $\Gamma_1 \sim 100$, combined with a larger zone with $R_2 \sim 10^{16}$~cm, $B_2 \sim 10^{-3}$~G, and $\Gamma_2 \sim 100$, can reproduce the observed emission \citep{Khangulyan2023_twozone}. The model naturally explains the temporal coincidence of X-ray and TeV emission, as both zones are causally connected and respond to the same underlying energy injection from the central engine.

The two-zone model predicts a specific spectral shape for the TeV emission. In the Thomson regime, the SSC spectrum follows:
\begin{equation}
E^2 \frac{dN}{dE} \propto E^{(3-p)/2}
\end{equation}
which, for $p \approx 2$--$2.5$, yields a flat to slightly rising spectrum ($E^2 dN/dE \propto E^{0.5}$ to $E^{0.25}$). This is consistent with the hard intrinsic spectrum inferred for GRB~221009A after EBL correction, especially when combined with a gradual transition to the KN regime at the highest energies. The model also predicts correlated variability between the X-ray and TeV bands, with possible time lags depending on the relative sizes and distances of the two emission regions.

{{Benchmark} 
 two‑zone SSC parameters.}  
To illustrate the two‑zone synchrotron self‑Compton scenario quantitatively, we summarise here the benchmark calculation of \citet{Khangulyan2023_twozone}.  In that work the emission region is divided into a compact “zone 1” with magnetic field \(B_{1}\) and a much larger “zone 2” with field \(B_{2}\).  Synchrotron photons produced in zone 1 provide the dominant target for inverse‑Compton scattering by electrons in zone 2.  The authors adopt \(B_{1}=1~\mathrm{G}\) and \(B_{2}=10^{-3}~\mathrm{G}\), with a common linear size \(R\simeq 10^{16}~\mathrm{cm}\), and inject a non‑thermal electron population with power‑law index \(\alpha=2.2\) and total power \(L_{0}\approx10^{39}~\mathrm{erg\,s^{-1}}\) split between the two zones in the ratio \(\kappa_{1}:\kappa_{2}=9:1\).  The energy density of synchrotron photons from zone 1 is then \(w_{\rm ph}\approx4\times10^{-5}\,\kappa_{1}\,(R/3\times10^{16}\,\mathrm{cm})^{-2}\,\mathrm{erg\,cm^{-3}}\), corresponding to an equivalent magnetic field \(B_{\rm eq}\approx3\times10^{-2}\,\kappa_{1}^{1/2}\,(R/3\times10^{16}\,\mathrm{cm})^{-1}\,\mathrm{G}\) in zone 2.  Assuming an acceleration efficiency \(\eta_{\rm acc}=10^{2}\), the maximum electron energy in zone 1 is \(E_{\rm cut,1}\simeq 6\,(\eta_{\rm acc}/10^{2})^{-1/2}\,(B_{1}/1\,\mathrm{G})^{-1/2}\,\mathrm{TeV}\), while in zone 2 it reaches \(\sim200\ \mathrm{TeV}\).  The resulting SSC spectrum forms a power‑law from a few GeV up to \(\sim10~\mathrm{TeV}\) with photon index \(\alpha_{\rm pred}=(\alpha+1)/2\approx1.6\) and then softens above \(\sim10~\mathrm{TeV}\) due to Klein–Nishina effects.  Even with these extreme parameters the predicted emission does not reproduce the hard intrinsic spectrum and 13 TeV photons observed by LHAASO, indicating that additional ingredients—such as time‑dependent microphysics, a structured jet, or a wind‑like external medium—are needed to match the data.

\subsubsection{Reverse Shock Emission}

The temporal coincidence of the TeV emission with the transition from prompt to afterglow-dominated phases suggests a possible connection to the reverse shock—a short-lived shock that propagates back into the GRB ejecta as it interacts with the surrounding medium. The reverse shock is expected to be most active during this transition phase and could potentially accelerate particles to higher energies than the forward shock due to its different physical conditions.

A comprehensive model has been developed in which the multi-TeV emission originates from the reverse shock region \citep{Zhang2023_reverseShock}. In this framework, the reverse shock efficiently accelerates both electrons and protons to ultra-relativistic energies. The TeV photons are then produced either through proton synchrotron radiation or through external inverse Compton scattering, where relativistic electrons in the reverse shock upscatter lower-energy photons from the forward shock or prompt emission.

\citet{Zhang2023_reverseShock} specifically analyze GRB 221009A and its TeV emission detected by LHAASO, dividing their analysis into two episodes:
\begin{itemize}
\item Episode I: Where external inverse-Compton (EIC) dominates due to strong prompt emission
\item Episode II: Where proton synchrotron becomes more prominent as prompt emission fades
\end{itemize}

This temporal evolution is important for understanding the changing emission mechanisms throughout the burst.

In the reverse shock model, the shell of GRB ejecta with initial Lorentz factor $\Gamma_0$ and width $\Delta$ begins to decelerate at the radius:
\begin{equation}
R_{\text{dec}} \approx \left(\frac{3E}{4\pi n m_p c^2 \Gamma_0^2}\right)^{1/3}
\end{equation}
where $E$ is the isotropic-equivalent energy, $n$ is the ambient density, and $m_p$ is the proton mass. At this radius, a reverse shock forms and propagates back into the ejecta, heating and decelerating it. The strength of the reverse shock is characterized by a dimensionless parameter:
\begin{equation}
\xi \approx \left(\frac{\Gamma_0^2 n}{n'}\right)^{1/4}
\end{equation}
where $n'$ is the comoving number density of the ejecta. For $\xi \gg 1$, the reverse shock is relativistic in the ejecta frame and can efficiently accelerate particles to high energies.

The magnetic field in the reverse shock region is expected to be much stronger than in the forward shock, especially if the GRB ejecta carries a significant magnetic field. The comoving magnetic field strength in the reverse shock can be estimated as:
\begin{equation}
B'_{\text{RS}} \approx \sqrt{32\pi \epsilon_B n m_p c^2 \Gamma_{\text{sh}}^2}
\end{equation}
where $\epsilon_B$ is the fraction of shock energy in magnetic fields and $\Gamma_{\text{sh}}$ is the Lorentz factor of the reverse shock in the ejecta frame. For typical parameters of GRB~221009A, with $\Gamma_0 \sim 1000$, $E \sim 10^{55}$ erg, and $n \sim 0.1$--$1$ cm$^{-3}$, the estimated $B'_{\text{RS}} \sim 10$--$100$ G \citep{Zhang2023_reverseShock}, much larger than the typical forward shock magnetic field of $\sim 0.1$--$1$ G.

The proton synchrotron scenario is particularly interesting, as it can naturally produce a hard spectrum extending to very high energies. Unlike electrons, protons have a much lower synchrotron cooling rate due to their larger mass, allowing them to reach higher energies before cooling becomes significant. The maximum energy of protons accelerated in the reverse shock is determined by the balance between acceleration and the dominant cooling or escape processes:
\begin{equation}
E_{p,\max} = \min\left[E_{p,\text{sync}}, E_{p,\text{dyn}}, E_{p,\text{adia}}\right]
\end{equation}
where $E_{p,\text{sync}}$ is the energy at which synchrotron cooling limits further acceleration, $E_{p,\text{dyn}}$ is the energy limited by the size of the acceleration region, and $E_{p,\text{adia}}$ is the energy limited by adiabatic cooling.

For a reverse shock with magnetic field $B' \sim 100$~G in the comoving frame and bulk Lorentz factor $\Gamma \sim 100$, protons can be accelerated to energies $E_p \sim 10^{18}$~eV, sufficient to produce synchrotron photons up to $\sim$100~TeV in the observer frame. The characteristic energy of synchrotron photons emitted by protons with energy $E_p$ is:
\begin{equation}
E_{\gamma,\text{sync}} \approx \Gamma \frac{3 h e B'}{4\pi m_p c} \left(\frac{E_p}{m_p c^2}\right)^2 \approx 10 \, \text{TeV} \left(\frac{\Gamma}{100}\right) \left(\frac{B'}{100 \, \text{G}}\right) \left(\frac{E_p}{10^{18} \, \text{eV}}\right)^2
\end{equation}

The resulting proton synchrotron spectrum follows:
\begin{equation}
E^2 \frac{dN}{dE} \propto E^{(3-p_p)/2}
\end{equation}
where $p_p$ is the power-law index of the proton energy distribution. For $p_p \approx 2$, this yields a spectral index of $(3-p_p)/2 = 1/2$, which is harder than the spectral index $\sim 0.2$ inferred from Fermi-LAT observations in Episode I \citep{Zhang2023_reverseShock}. This harder spectral index enhances the fraction of O(10 TeV) photons despite EBL attenuation, which is consistent with the hard intrinsic spectrum inferred for GRB~221009A.

The reverse shock model naturally explains the timing of the TeV emission, which coincides with the transition from prompt to afterglow-dominated emission, when the reverse shock is expected to be most active. It also predicts a relatively short duration for the TeV emission, consistent with the LHAASO observations, as the reverse shock is a transient phenomenon that quickly fades as the ejecta decelerates.

\citet{Zhang2023_reverseShock} also consider an alternative scenario where the TeV emission results from the external inverse Compton (EIC) process, with electrons in the reverse shock upscattering photons from the forward shock or prompt emission. This process can also produce TeV photons, but faces challenges in explaining the observed hard spectrum and high flux due to KN suppression and the relatively low density of target photons. The model suggests that EIC dominates when $\epsilon_e > \epsilon_B$ (Episode I), while proton synchrotron becomes more important when $\epsilon_B$ increases (Episode II).

An important advantage of the reverse shock proton synchrotron model is that it can naturally avoid the internal gamma-gamma absorption that potentially affects electron-based models. Since the proton synchrotron emission directly produces TeV photons without requiring a dense field of lower-energy target photons (as in SSC or EIC models), the opacity for internal absorption is significantly reduced.

The reverse shock model also predicts associated neutrino production, which could be an important multi-messenger aspect of these events. As \citet{Zhang2023_reverseShock} discuss, the neutrino energy spectrum predicted in Episode I has two bumps, with the low-energy bump at ~PeV energies due to photomeson production between high-energy protons and prompt target photons, and the high-energy bump at ~10 EeV resulting from interactions between UHE protons and lower-energy synchrotron photons from the reverse shock.

\subsubsection{Pair‑Balance Synchrotron Self‑Compton Emission}
{

In the pair‑balance scenario, high‑energy inverse‑Compton photons collide with synchrotron photons ahead of the shock and generate copious electron–positron pairs.  The freshly created pairs carry most of the energy injected by the shock and regulate the radiative efficiency: the Compton parameter $Y\equiv U_{\rm ph}/U_B$ saturates at order unity, and the SSC peak energy stabilises near $E_{\gamma,\rm SSC}\sim1$~TeV.  Consequently, the SSC and synchrotron components have comparable luminosities and the multi‑TeV flux becomes largely insensitive to $\varepsilon_e$ and $\varepsilon_B$ \citep{Derishev2024_GRB221009A}.  Derishev \& Piran modelled the contemporaneous phase of GRB 221009A with a two‑element blast wave and an intermittent energy supply, finding that a very narrow jet with half‑opening angle $\theta_j\simeq0.07^{\circ}(500/\Gamma_0)$ propagating into a wind‑like medium reproduces the entire TeV light curve except for three peaks that coincide with prompt pulses \citep{Derishev2024_GRB221009A}.  In their best‑fit solution photon–photon annihilation is negligible and the pair‑balance condition naturally explains the hard TeV spectrum detected by LHAASO.  This makes pair‑balance SSC emission a promising alternative to conventional afterglow models for explaining the multi‑TeV radiation from the BOAT.}

\subsubsection{Stochastic Turbulent Acceleration and a Second Electron Component}
{An alternative explanation invokes downstream turbulence to stochastically re‑accelerate electrons.  Gong et~al. argued that the hard multi‑TeV spectrum of GRB 221009A cannot be accounted for by the standard one‑zone SSC model and introduced a second component of electrons accelerated via transit‑time damping in magnetic turbulence behind the shock \citep{Gong2025_Turbulence}.  They solved the coupled evolution of particles and turbulence with a Fokker–Planck approach and showed that the additional electron population acquires a hard spectrum whose inverse‑Compton emission hardens the intrinsic spectrum above $\sim10$ TeV.  Their two‑zone model, which adopts a narrow jet ($\theta_j\approx0.8^{\circ}$) and a wind‑like circumburst medium, reproduces the LHAASO light curve and spectrum without overproducing the emission at lower energies \citep{Gong2025_Turbulence}.  In this framework the initial magnetic field and turbulence spectrum play only a minor role; the turbulent component is negligible at early times but becomes dominant as the blast wave decelerates and the peak energy of the re‑accelerated electrons shifts into the TeV band.  This stochastic acceleration scenario thus provides another viable mechanism for the multi‑TeV emission observed in GRB~221009A.

}

\subsection{Challenge II: Propagation of Multi-TeV Gamma Rays}

The second fundamental challenge concerns the propagation of multi-TeV gamma rays from GRB~221009A to Earth despite the expected severe attenuation by the EBL. For a source at redshift $z=0.151$, standard EBL models predict an optical depth $\tau \approx 10$ at $E \approx 10$~TeV, corresponding to a flux attenuation factor of $e^{-10} \approx 4.5 \times 10^{-5}$ \cite{Abdalla2024_void,Finke2023_LIV}. 
Several explanations have been proposed to address this propagation puzzle, ranging from astrophysical effects related to the inhomogeneous distribution of the EBL to more exotic scenarios involving new physics beyond the Standard Model.

\subsubsection{Reduced EBL Density Due to Cosmic Voids}

One of the more conservative explanations for the detection of multi-TeV photons from GRB~221009A involves the possibility that the line of sight traverses regions of lower-than-average EBL density, known as cosmic voids. The universe's large-scale structure is characterized by a cosmic web of filaments, clusters, and voids—the latter being vast regions of significantly lower-than-average matter density. These cosmic voids, spanning tens to hundreds of megaparsecs, naturally contain fewer galaxies and stars than average regions, resulting in locally reduced EBL density \citep{Abdalla2017_EBL, Kudoda2017_voids, Abdalla2024_void}.

The potential impact of cosmic voids on gamma-ray propagation was recognized well before GRB~221009A. Earlier studies demonstrated that the presence of voids along the line of sight to distant gamma-ray sources could significantly reduce the effective optical depth for VHE photons \citep{Abdalla2017_EBL}. This effect is particularly relevant for TeV astronomy, as it could extend the observable universe at the highest energies and potentially explain anomalous spectral hardening observed in some blazars.

In the specific context of GRB~221009A, our detailed analysis \citep{Abdalla2024_void} modeled the impact of cosmic voids on the EBL density along the line of sight and calculated the resulting reduction in gamma-gamma opacity. This analysis considered voids of various sizes and density contrasts (the ratio of void density to mean cosmic density) ranging from 0.1 to 0.5, spanning from small voids of a few tens of megaparsecs to larger structures of up to 250~megaparsecs.

Using this framework, we found that the $\gamma\gamma \to e^+e^-$ opacity for VHE gamma rays can be reduced by approximately 10\% for intervening cosmic voids along the line of sight with a combined radius of 110~Mpc (typically found from void catalogues), and up to 30\% at around 13~TeV (the highest-energy photon confidently detected from GRB~221009A) for voids with a combined radius of 250~Mpc \citep{Abdalla2024_void}. This reduction is substantially higher for TeV photons compared to GeV photons, attributable to the broader target photon spectrum that TeV photons interact with.

The deficit due to cosmic voids is more pronounced for TeV photons than for GeV photons because TeV photons interact with a broader range of target photons in the EBL spectrum, spanning from infrared to far-infrared wavelengths, which are more affected by the reduced star formation in voids. In contrast, GeV photons primarily interact with optical and near-infrared photons, which have a more uniform distribution across cosmic structures. This energy-dependent effect could potentially explain why the observed spectrum of GRB~221009A appears to harden at the highest energies after correcting for standard EBL absorption models.

To quantify the statistical significance of the void scenario, we performed Monte Carlo simulations, generating synthetic observations of GRB~221009A under various assumptions about the void distribution along the line of sight \citep{Abdalla2024_void}. Our analysis found that the presence of voids can increase the probability of detecting the observed number of multi-TeV photons by up to two orders of magnitude compared to a homogeneous EBL distribution. However, even with the most optimistic void configurations, this probability remains low ($\sim 10^{-4}$), suggesting that additional factors may be at play.

The void scenario has several attractive features: it requires no new physics beyond standard cosmology, is consistent with our understanding of cosmic structure formation, and can be tested with independent observations of the matter distribution along the line of sight to GRB~221009A. However, it faces challenges in explaining the full magnitude of the observed effect, as even the most optimistic void configurations cannot reduce the optical depth by more than a factor of $\sim$2, whereas the observations suggest a required reduction by a factor of $\sim$10 or more \citep{Abdalla2024_void, Finke2023_LIV}. This indicates that while voids may contribute to the solution, they are unlikely to fully resolve the puzzle of GRB~221009A's multi-TeV~emission.

Further support for the astrophysical relevance of cosmic voids in high-energy gamma-ray propagation comes from broader population studies. \citet{Furniss2025_voidiness} investigated the statistical alignment of gamma ray-emitting active galactic nuclei (AGN) with underdense regions in the large-scale structure and found that, particularly at higher redshifts, these sources are more likely to lie behind void-rich sightlines compared to radio-quiet quasars. This result complements our GRB~221009A findings by indicating that cosmic voids may not only reduce gamma-ray attenuation due to locally diminished EBL densities but may also preferentially host or reveal populations of sources whose emission is otherwise suppressed in denser environments.

It is worth noting that the sky region encompassing GRB~221009A, positioned at \mbox{RA = 288.264°} and Dec = 19.768°, is not included in current void catalogs derived from the Sloan Digital Sky Survey SDSS DR7. Consequently, accurately estimating the extent of voids along our line of sight to GRB~221009A remains challenging. Future spectroscopic surveys with wider sky coverage may provide more definitive constraints on the void distribution in this direction, allowing for more precise estimates of the EBL reduction~effect.

\begin{figure}[H] 

 \includegraphics[width=\columnwidth, angle=0]{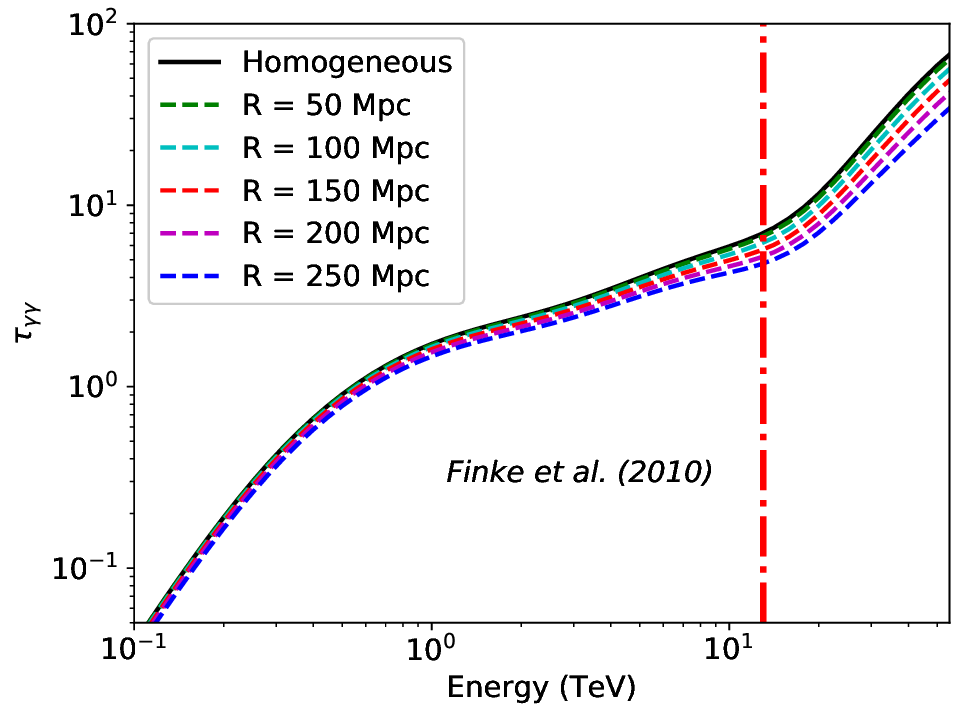}\\
 \includegraphics[width=\columnwidth, angle=0]{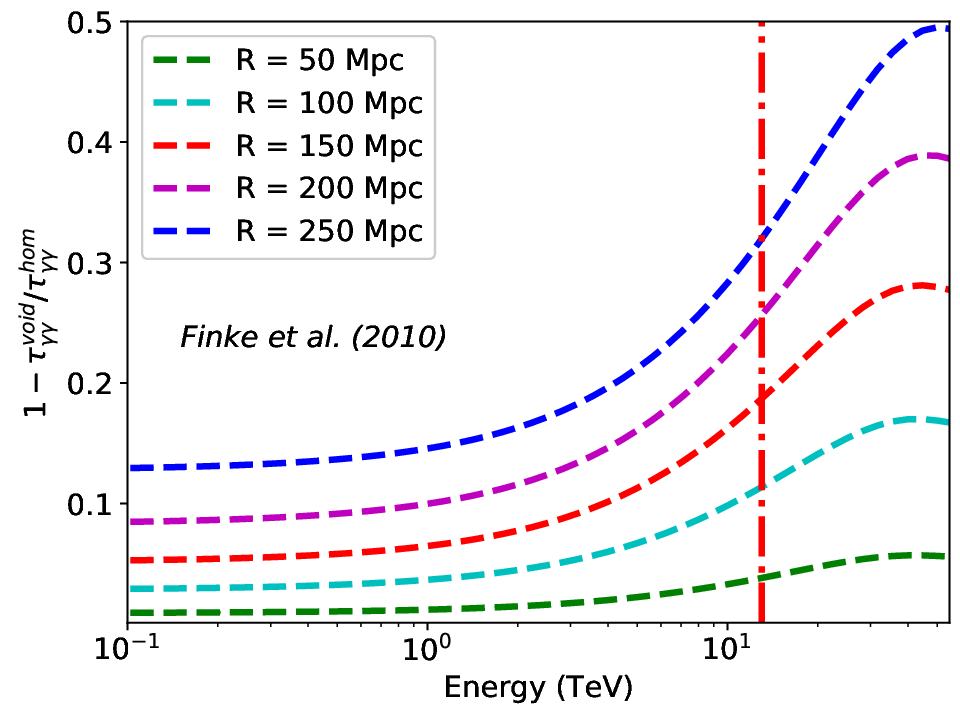}
 \caption{{ {Top panel:} 
} $\gamma \gamma \to e^+e^-$ optical depth due to the EBL as a function of the $\gamma$-ray energy for a source located at redshift $z = 0.151$. The black solid line indicates the optical depth for a homogeneous EBL distribution. The dashed lines show the opacities in the presence of voids of various sizes, as indicated by the legend. The vertical red dot-dashed line in both panels indicates a photon energy of 13~TeV.
 { Bottom panel:} Relative deficit of the optical depth with and without voids as a function of the $\gamma$-ray energy for the same cases as in the top panel. Figure adapted from \cite{Abdalla2024_void}.
\label{opaf}}
\end{figure}

{

Abdalla et~al.\ showed that intervening cosmic voids can reduce the $\gamma\gamma$ optical depth by at most $\sim30\%$ at $\sim 13\,\mathrm{TeV}$, see Figure \ref{opaf}.  Even with such a reduction, the implied multi‑TeV luminosity remains extreme.  LHAASO measured a peak energy flux of \linebreak  $\nu F_\nu \approx 1.2\times10^{-5}\,\mathrm{erg\,cm^{-2}\,s^{-1}}$ in the $0.3$–$5$ TeV band. For GRB~221009A at $z=0.151$ (luminosity distance $D_L\simeq745\ \mathrm{Mpc}$), this corresponds to an isotropic‑equivalent luminosity 
\[
L_{\gamma,\mathrm{TeV}}\simeq 4\pi D_L^2\,\nu F_\nu \approx 7.3\times 10^{50}\ \mathrm{erg\,s^{-1}},
\]
consistent with the value reported by the LHAASO collaboration \citep{Cao2023a}.  Even if $\tau_{\gamma\gamma}$ were reduced by 30\%, the intrinsic multi‑TeV luminosity would still be of order $10^{50}$–$10^{51}\ \mathrm{erg\,s^{-1}}$, which greatly exceeds typical afterglow energetics unless the emission is highly collimated. In summary, while cosmic voids offer a conservative astrophysical explanation that partially accounts for the detection of multi-TeV photons from GRB~221009A, the magnitude of the effect appears insufficient to fully explain the observations.
}

\subsubsection{New Physics: Axion-Like Particles}

Axion-like particles (ALPs) offer a compelling explanation for the observed multi-TeV emission from GRB~221009A. These hypothetical light pseudoscalar particles couple to photons in the presence of magnetic fields, enabling oscillations between the two states—analogous to neutrino flavor oscillations but involving conversion between different particle types \cite{Troitsky2022_ALP,Wang:2023okw}.

The theoretical foundation for ALPs emerges from various extensions of the Standard Model, including string theory and models addressing the strong CP problem in quantum chromodynamics \cite{Galanti2023_ALP,Nakagawa:2022wwm}. ALPs are characterized by two fundamental parameters: mass $m_a$ and coupling constant to photons $g_{a\gamma}$. According to Troitsky \cite{Troitsky2022_ALP} and Rojas et al. \cite{Rojas:2023jdd}, the phenomenologically relevant parameter space for astrophysical observations spans $m_a \sim 10^{-11}$--$10^{-7}$~eV and $g_{a\gamma} \sim 10^{-12}$--$10^{-10}$~GeV$^{-1}$.

The ALP-photon interaction is governed by the Lagrangian term as described in Troitsky \cite{Troitsky2022_ALP} and Wang and Ma \cite{Wang:2023okw}:
\begin{equation}
\mathcal{L}_{a\gamma} = -\frac{1}{4}g_{a\gamma}aF_{\mu\nu}\tilde{F}^{\mu\nu}
\end{equation}
where $a$ represents the ALP field, $F_{\mu\nu}$ denotes the electromagnetic field tensor, and $\tilde{F}^{\mu\nu}$ is its dual. This coupling enables photon-ALP oscillations in external magnetic fields, with a conversion probability in a homogeneous field given by Troitsky \cite{Troitsky2022_ALP}:
\begin{equation}
P(\gamma \rightarrow a) = \left(\frac{g_{a\gamma}BL}{2}\right)^2 \times \frac{\sin^2(\Delta L/2)}{(\Delta L/2)^2}
\end{equation}
where $B$ represents the transverse magnetic field component, $L$ denotes the propagation distance, and $\Delta$ corresponds to the momentum difference between photon and ALP states.

In the context of GRB~221009A, Galanti et al. \cite{Galanti2023_ALP} and Troitsky \cite{Troitsky2023uwu} propose that gamma rays produced in the GRB jet can oscillate into ALPs before encountering the EBL. Since ALPs do not interact electromagnetically, they traverse cosmological distances unimpeded before reconverting to gamma rays in the Milky Way's magnetic field, effectively circumventing EBL absorption as demonstrated by Wang and Ma \cite{Wang:2023okw} and Rojas et al. \cite{Rojas:2023jdd}.

The complete photon-ALP system propagation is described by a mixing matrix formulated by Wang and Ma \cite{Wang:2023okw}:
\begin{equation}
M = \begin{pmatrix}
\Delta_{\text{pl}} + \Delta_{\text{QED}} + \Delta_{\text{abs}} & 0 & \Delta_{a\gamma} \\
0 & \Delta_{\text{pl}} + \Delta_{\text{abs}} & 0 \\
\Delta_{a\gamma} & 0 & \Delta_a
\end{pmatrix}
\end{equation}
where the $\Delta$ terms represent different contributions to the propagation: plasma effects, QED vacuum polarization, absorption, ALP-photon mixing, and ALP mass.

Troitsky \cite{Troitsky2022_ALP} demonstrates that this mechanism could potentially explain the observed multi-TeV emission from GRB~221009A for ALP masses in the range $m_a \approx 10^{-11}$--$10^{-7}$~eV and photon-ALP coupling constants $g_{a\gamma} \approx (3-5) \times 10^{-12}$~GeV$^{-1}$. This finding is corroborated by Galanti et al. \cite{Galanti2023_ALP}, who arrive at similar parameter constraints through independent analysis. Detailed calculations of the expected photon-ALP conversion probability along the full path from GRB~221009A to Earth have been performed by Troitsky \cite{Troitsky2023uwu} and Wang and Ma \cite{Wang:2023okw}, accounting for magnetic fields in the host galaxy, intergalactic medium, and Milky Way.

According to Troitsky \cite{Troitsky2022_ALP}, ALP-photon mixing primarily in the Milky Way, rather than in intergalactic space, may be crucial for explaining the observations. This conclusion was further refined in subsequent work by the same author \cite{Troitsky2023uwu}, which developed a more detailed model of photon-ALP conversion in the host galaxy of GRB~221009A, demonstrating that conversion in both the host galaxy and the Milky Way is necessary for a comprehensive explanation.

The ALP scenario provides a natural explanation for the energy-dependent transparency of the universe to gamma rays, as the photon-ALP conversion probability depends on photon energy, as shown by Rojas et al. \cite{Rojas:2023jdd} and Wang and Ma \cite{Wang:2023okw}. This model also predicts specific spectral and polarization signatures that could be tested with future observations. Dzhatdoev et al. \cite{Dzhatdoev17} note that the ALP-induced spectral modulations would depend on the detailed magnetic field configuration along the line of sight, potentially creating distinctive features in the energy spectrum.

Significantly, Galanti et al. \cite{Galanti2023_ALP} and Troitsky \cite{Troitsky2022_ALP} demonstrate that while LIV can explain the possible 251 TeV photon reported by the Carpet-2 detector, it cannot simultaneously account for the LHAASO observations. In contrast, the ALP model with parameters $m_a \approx (10^{-11}$--$10^{-7})$~eV and $g_{a\gamma} \approx (3-5) \times 10^{-12}$~GeV$^{-1}$ can explain both observations, as independently confirmed by Troitsky \cite{Troitsky2022_ALP} and Rojas et al. \cite{Rojas:2023jdd}.

Rojas et al. \cite{Rojas:2023jdd} analyze the spectral signatures of GRB~221009A based on ALP candidates, demonstrating that converting TeV photons into ALPs at the host galaxy, avoiding EBL absorption, and then reconverting them into photons within the Milky Way relaxes the requirement of very high photon fluxes at the source. Their analysis incorporates the energy dependence of the survival probability to determine the spectral conditions required for ALP injection.

Nakagawa et al. \cite{Nakagawa:2022wwm} establish a connection between the ALP explanation of GRB~221009A and dark matter models, proposing that ALPs could be produced in the early universe through first-order phase transitions and could constitute a significant fraction of dark matter. This provides additional theoretical motivation for the existence of ALPs in the parameter range relevant for explaining GRB~221009A.

Dzhatdoev et al. \cite{Dzhatdoev17}, while focusing on blazars rather than GRBs, provide important methodological insights for distinguishing genuine ALP effects from electromagnetic cascades that can sometimes mimic the spectral features attributed to ALPs. Their work offers analytical techniques to differentiate between these scenarios, which may prove valuable for future observations of GRBs.

Research by Galanti et al. \cite{Galanti2023_ALP} and Troitsky \cite{Troitsky2023uwu} suggests that the introduction of ALPs in magnetized media provides a compelling explanation for the detection of GRB~221009A's multi-TeV emission, potentially offering observational evidence for physics beyond the Standard Model. This interpretation gains further support from the fact that ALPs are a generic prediction of many theories extending the Standard Model toward a more unified framework, as noted by Nakagawa et al. \cite{Nakagawa:2022wwm}, and that they could simultaneously address the dark matter problem, as proposed by Nakagawa et al. \cite{Nakagawa:2022wwm} and Troitsky \cite{Troitsky2022_ALP}.

{
The photon–ALP mixing model proposed for GRB 221009A adopts very light ALPs with $m_a\sim 10^{-10}\,\mathrm{eV}$ and a coupling $g_{a\gamma\gamma}\approx (3$--$5)\times 10^{-12}\,\mathrm{GeV}^{-1}$ in order to maximise the conversion probability at $E\sim15\,$TeV and reproduce the deabsorbed LHAASO spectrum\citep{Galanti2023_ALP}.  It is important to note that these parameters should be checked against existing laboratory and astrophysical constraints.  The CAST solar axion search has recently reported an upper limit $g_{a\gamma\gamma}\lesssim 5.8\times 10^{-11}\,\mathrm{GeV}^{-1}$ (95\% C.L.) for $m_a\lesssim 0.02\,\mathrm{eV}$\citep{CAST2024}, so the couplings invoked here lie below the latest CAST bound.  However, more stringent astrophysical limits come from the Chandra X‑ray study of the quasar H1821+643, which excludes $g_{a\gamma\gamma} > 6.3\times 10^{-13}\,\mathrm{GeV}^{-1}$ for $m_a<10^{-12}\,\mathrm{eV}$ at 99.7\% confidence\citep{SiskReynes2021}.  The favoured range $(3$--$5)\times 10^{-12}\,\mathrm{GeV}^{-1}$ therefore approaches or exceeds the most stringent astrophysical limits, and this tension should be acknowledged when discussing ALP interpretations.

}

\subsubsection{New Physics: Lorentz Invariance Violation}

Lorentz invariance—the principle that the laws of physics are the same for all observers moving at constant velocities relative to each other—is a cornerstone of modern physics, underlying both special relativity and quantum field theory. However, several quantum gravity theories suggest that this symmetry might be broken at very high energies, approaching the Planck scale ($E_{\text{Pl}} \approx 1.22 \times 10^{19}$ GeV). This phenomenon, known as Lorentz invariance violation (LIV), could have observable consequences for the propagation of high-energy particles over cosmological distances \citep{Coleman1999_LIV}.

LIV has been proposed as a potential explanation for the observed multi-TeV emission from GRB~221009A \citep{Baktash2023_LIV,Finke2023_LIV,LHAASO2024_LIV,Yang2024_LIV}. In quantum gravity theories, Lorentz symmetry may be broken at energies approaching the Planck scale, leading to modifications in the dispersion relation of photons and other particles. These modifications can affect the kinematics of pair production, potentially allowing high-energy gamma rays to travel further than expected in standard physics.

The modified dispersion relation for photons can be written as \citep{Abdalla2018, Abdalla2024_LIV}:
\begin{equation}
E^2 = p^2c^2 \left[1 - \sum_{n=1}^{\infty} S \left(\frac{E}{E_{\text{QG},n}}\right)^n\right],
\label{eq:liv_dispersion}
\end{equation}
where $c$ is the conventional speed of light in vacuum, $S = +1$ and $S = -1$ refer to the subluminal and superluminal scenarios, respectively, and $E_{\text{QG},n}$ represents the quantum gravity energy scale. At this scale, the Compton wavelength and Schwarzschild radius of a particle with mass $m$ are of the same order of magnitude, and the effects of quantum gravity cannot be ignored \citep{Yang2024_LIV}. The Planck mass is estimated as $M_{\text{Pl}} = \sqrt{\hbar c/G} \approx 1.22 \times 10^{19}$ GeV, which is significantly higher than that achievable in collider experiments or involved in currently known astrophysical processes.

In this analysis, we only consider the case where the lowest order term is $n = 1$ or $n = 2$. The group velocity of light in vacuum can be deduced from the above dispersion relationship as \citep{Yang2024_LIV}:
\begin{equation}
v_g(E) = \frac{\partial E}{\partial p} \approx c \left[1 - S \frac{n+1}{2} \left(\frac{E}{E_{\text{QG},n}}\right)^n\right].
\label{eq:group_velocity}
\end{equation}

The time lag between photons of different energies due to LIV is given by \citep{Yang2024_LIV}:
\begin{equation}
\Delta t_{\text{LIV}}(n, E_2, E_1) \approx S \frac{n+1}{2} \frac{E_2^n - E_1^n}{E_{\text{QG},n}} \int_0^z \frac{(1+z')^n}{H(z')} dz',
\label{eq:time_lag}
\end{equation}
where $H(z) = H_0 \sqrt{\Omega_m(1+z)^3 + \Omega_\Lambda}$ is the Hubble parameter at redshift $z$. For numerical convenience, we can define two additional parameters \citep{Yang2024_LIV}: $\eta_1 = S M_{\text{Pl}}/E_{\text{QG},1}$ and $\eta_2 = 10^{-16} S (M_{\text{Pl}}/E_{\text{QG},2})^2$.

For $n=1$, the time lag simplifies to:
\begin{equation}
\Delta t_{\text{LIV}}(1, E_2, E_1) = 5.87\eta_1 \left(\frac{E_2}{\text{TeV}} - \frac{E_1}{\text{TeV}}\right) \text{ s},
\label{eq:time_lag_n1}
\end{equation}
and for $n=2$:
\begin{equation}
\Delta t_{\text{LIV}}(2, E_2, E_1) = 7.77\eta_2 \left[\left(\frac{E_2}{\text{TeV}}\right)^2 - \left(\frac{E_1}{\text{TeV}}\right)^2\right] \text{ s}.
\label{eq:time_lag_n2}
\end{equation}

For the case of GRB~221009A, the subluminal scenario ($S = +1$) is of particular interest, as it can lead to an increase in the threshold energy for pair production, potentially allowing multi-TeV photons to propagate further than expected in conventional physics. The LHAASO Collaboration has conducted stringent tests of LIV using GRB~221009A observations \citep{LHAASO2024_LIV}. Through the maximum likelihood method, Yang et al. \citep{Yang2024_LIV} determined the 95\% confidence level lower limits to be:
\begin{equation}
E_{\text{QG},1} > 14.7(6.5) \times 10^{19} \text{ GeV}
\label{eq:limit_n1}
\end{equation}
for the subluminal (superluminal) scenario of $n = 1$, and
\begin{equation}
E_{\text{QG},2} > 12.0(7.2) \times 10^{11} \text{ GeV}
\label{eq:limit_n2}
\end{equation}
for the subluminal (superluminal) scenario of $n = 2$.

Similarly, Baktash et al. \citep{Baktash2023_LIV} found that for a linear ($n=1$) dependence of the photon velocity on energy, the lower limit on the subluminal (superluminal) LIV scale is approximately $5.9$ ($6.2$) $M_{\text{Pl}}$. For a quadratic model ($n=2$), the limits are \mbox{$5.8$ ($4.6$) $\times 10^{-8}$ $M_{\text{Pl}}$}. These results are comparable to the stringent limits obtained from other sources and, as independent bounds obtained from a different redshift, confirm their robustness.

The appeal of the LIV explanation lies in its ability to simultaneously address both the detection of multi-TeV photons from GRB~221009A and the hard intrinsic spectrum inferred after standard EBL correction. In the LIV scenario, the effective optical depth for pair production is reduced at the highest energies, leading to less attenuation and a harder observed spectrum. This naturally explains why the spectral anomaly becomes most pronounced above $\sim$10 TeV.

Finke \citep{Finke2023_LIV} conducted a detailed analysis showing that standard EBL models predict an extremely low survival probability for multi-TeV photons from GRB~221009A. Specifically, the survival probability of $\exp[-\tau_{\gamma\gamma}(18 \text{ TeV})] < 4.5 \times 10^{-5}$ makes the detection of such high-energy photons highly improbable under conventional physics. For the quadratic ($n=2$) case, Finke found that $E_{\text{QG}} \lesssim 10^{-6} E_{\text{Planck}}$ (95\% confidence limits), suggesting that this could potentially be the first evidence for subluminal LIV.

However, the LIV scenario is not without challenges. The energy scale $E_{\text{QG},1}$ derived from GRB~221009A is in tension with constraints from other astrophysical observations, particularly those from previous GRBs. For instance, observations of GRB~090510 by the Fermi-LAT telescope have placed a lower limit of $E_{\text{QG},1} > 9.3 \times 10^{19}$~GeV for $n = 1$ \citep{Abdo2009_LIVlimit}, which is marginally inconsistent with some values required to explain GRB~221009A. This tension suggests that if LIV is responsible for the observed multi-TeV emission, it may require a more complex energy dependence than the simple power-law form assumed in most models.

The LHAASO Collaboration \citep{LHAASO2024_LIV} has conducted the most comprehensive tests of LIV using GRB~221009A observations to date. Their analysis utilizes both the Kilometer Square Array (KM2A) and Water Cherenkov Detector Array (WCDA) components of LHAASO, providing unprecedented sensitivity to high-energy gamma rays. By simultaneously applying pair view and maximum likelihood methods to analyze the temporal evolution of the afterglow, they achieved the most stringent constraints on the quantum gravity energy scale. Notably, for the subluminal scenario with $n=1$, their lower limit exceeds the Planck scale, challenging many quantum gravity models that predict stronger LIV effects.

Yang et al. \citep{Yang2024_LIV} demonstrated that the rapid rise and slow decay behaviors of the GRB~221009A afterglow impose particularly strong constraints on the subluminal scenario, while constraints are weaker for the superluminal scenario. Their analysis highlights the importance of employing sources with wide energy spectra in LIV studies, as the difference between photon arrival times increases with energy difference. GRB~221009A's exceptional brightness and high-energy emission make it uniquely valuable for these tests, serving as an extraordinary laboratory for probing fundamental physics.

Another challenge for the LIV scenario is that it predicts a specific energy dependence for the gamma-ray horizon—the maximum distance at which sources of a given energy can be detected. This prediction could potentially be tested with observations of other TeV sources at various redshifts, providing a crucial test of the LIV hypothesis. However, current observational constraints from blazars and other TeV sources are not yet sufficient to definitively rule out or confirm the LIV explanation for GRB~221009A.

It is worth noting that LIV is not the only proposed explanation for the multi-TeV emission from GRB~221009A. Alternative mechanisms include photon conversion to ALPs in magnetic fields, reduced EBL density due to cosmic voids along the line of sight, and intrinsic spectral properties of the GRB itself. Each of these explanations has its own strengths and challenges, and the current observational data does not definitively favor one over the others. Future observations of high-energy GRBs, particularly with next-generation gamma-ray observatories, will be crucial for distinguishing between these competing explanations.

In summary, while LIV provides a theoretically motivated explanation for the observed multi-TeV emission from GRB~221009A, the constraints derived from this exceptional event are pushing the boundaries of quantum gravity theories. The fact that some limits exceed the Planck scale suggests either that LIV effects are extremely weak—perhaps too weak to explain the observations—or that the simple power-law parameterization commonly used may not capture the full complexity of potential quantum gravity effects on photon propagation.

\subsubsection{New Physics: Heavy Sterile Neutrino Decay}

The observed multi-TeV emission from GRB~221009A presents a significant challenge to conventional astrophysical models, particularly due to the expected attenuation of high-energy photons by the EBL. A compelling theoretical framework that addresses this challenge involves the decay of heavy sterile neutrinos—hypothetical particles that do not participate in standard model weak interactions but can mix with active neutrinos through mass terms.

Sterile neutrinos with masses in the range $m_N \sim 10$--$100$ GeV could be produced in the GRB environment through various mechanisms, including hadronic processes in the jet \cite{Brdar2023_heavyNu, Smirnov2023_heavyNu, Huang2023_heavyNu}. The production efficiency depends critically on the mixing between sterile and active neutrino states, particularly in the muon flavor sector as proposed by Brdar et al. \cite{Brdar2023_heavyNu}. These heavy neutrinos, once produced, would propagate essentially unimpeded across cosmological distances before decaying via channels such as $N \rightarrow \nu \gamma$ or $N \rightarrow \nu \gamma \gamma$, generating high-energy photons at locations significantly closer to Earth than the GRB~source.

This scenario elegantly circumvents the EBL attenuation problem that plagues conventional interpretations. The heavy sterile neutrinos, being weakly interacting, can traverse the intergalactic medium without significant interaction. The decay photons subsequently produced would only need to travel a fraction of the total distance from the GRB to Earth, thereby experiencing substantially reduced optical depth. Quantitatively, for an 18 TeV photon originating directly from GRB~221009A at redshift $z \approx 0.15$, the survival probability would be approximately $\exp(-15) \approx 10^{-7}$ \cite{Huang2023_heavyNu}, making direct observation exceedingly improbable within standard physics frameworks.

Detailed modeling by Huang et al. \cite{Huang2023_heavyNu} demonstrates that a sterile neutrino with mass $m_N \approx 50$ GeV and mixing parameter $\sin^2 \theta \approx 10^{-8}$--$10^{-6}$ could produce a photon flux consistent with LHAASO observations. Their analysis incorporates both the production of heavy neutrinos in the GRB environment and their subsequent decay kinematics. Importantly, this parameter space remains compatible with existing terrestrial constraints on sterile neutrinos, as emphasized by both Brdar et al. \cite{Brdar2023_heavyNu} and Smirnov and Trautner \cite{Smirnov2023_heavyNu}.

The sterile neutrino decay scenario predicts distinctive observational signatures that could potentially be tested with future multi-messenger observations. The energy spectrum of decay photons would exhibit characteristic features determined by the sterile neutrino mass and decay channels. For instance, in the two-body decay $N \rightarrow \nu \gamma$, the photon energy in the laboratory frame depends on the sterile neutrino energy and mass, potentially creating identifiable spectral features. Additionally, the temporal profile of the emission would be influenced by the sterile neutrino lifetime, introducing time delays relative to the prompt emission that scale with the mixing parameters and neutrino mass.

Smirnov and Trautner \cite{Smirnov2023_heavyNu} further elaborate that for this mechanism to explain the observed high-energy gamma rays, two critical conditions must be satisfied: (i) the GRB active neutrino fluence must approach the observed limit, and (ii) the branching ratio for $N \rightarrow \nu\gamma$ must be at least of order 10\%. Their analysis indicates that a heavy neutrino with mass $m_N \sim 0.1$ MeV could satisfy these conditions while remaining consistent with current observational constraints.

The viability of the heavy sterile neutrino scenario hinges on several theoretical and observational considerations. First, the production efficiency of heavy neutrinos in the GRB jet depends on detailed assumptions about the jet composition and physical conditions, which remain uncertain. Second, the mixing between sterile and active neutrinos must be carefully balanced within a narrow parameter space: too large, and the sterile neutrinos would decay prematurely; too small, and insufficient numbers would be produced to explain the observed flux. Third, as highlighted by Brdar et al. \cite{Brdar2023_heavyNu}, the transition magnetic moment portal between active and sterile neutrinos plays a crucial role in determining the decay rate, with the optimal scenario involving both mixing and transition magnetic moment mechanisms.

Despite these challenges, the heavy sterile neutrino interpretation remains particularly compelling in light of the difficulties faced by conventional explanations. The scenario naturally accommodates the observed multi-TeV emission while remaining consistent with existing constraints from laboratory experiments and cosmological observations. Future observations of high-energy emission from GRBs, combined with improved constraints on sterile neutrinos from terrestrial experiments and cosmological probes, will be instrumental in testing this hypothesis and potentially revealing new physics beyond the Standard~Model.

\subsubsection{Ultrahigh-Energy Cosmic Ray Cascades}

The observed multi-TeV emission from GRB~221009A presents a profound challenge to conventional astrophysical models, particularly due to the expected severe attenuation of high-energy photons by the EBL. A compelling alternative explanation involves the production of secondary gamma rays through electromagnetic cascades initiated by ultra-high-energy cosmic rays (UHECRs) accelerated in the GRB environment. This mechanism provides a natural framework for understanding the anomalous TeV emission while circumventing the EBL attenuation {problem} 
 \cite{DasRazzaque2023_AA, Das2025_arXiv}.
GRB~221009A, with its unprecedented brightness and multi-wavelength coverage, offers a unique opportunity to investigate UHECR acceleration in GRBs. Located at redshift $z = 0.151$ (corresponding to a luminosity distance of 745 Mpc) \cite{Malesani2023_AA} and releasing an isotropic-equivalent radiation energy of $E_{\rm iso} \sim 10^{55}$ ergs \cite{DasRazzaque2023_AA, He2023_cascade}, this burst represents the brightest GRB ever detected. The detection of photons with energies extending up to $\sim$18 TeV by LHAASO \cite{Huang2022_ATel}, despite the expected severe attenuation by the EBL, strongly suggests the involvement of non-conventional physical processes.

In the UHECR cascade scenario, the fundamental mechanism involves several interconnected processes:

\begin{enumerate}
\item \textit{UHECR Acceleration}: Protons are accelerated to energies of $10^{18}$--$10^{20}$ eV in the GRB jet, particularly at the energy dissipation radius during the prompt emission phase. The energy dissipation radius of GRB~221009A is estimated to be \cite{He2023_cascade}:
\begin{equation}
R = 2\Gamma^2 ct_v/(1+z) = 10^{15} \text{ cm } (1+z)^{-1}\Gamma_2^2 \frac{t_{v,-2}}{1}
\end{equation}
where the variability timescale is measured as $t_v = 0.082$ s \cite{He2023_cascade}, $\Gamma$ is the bulk Lorentz factor, and $c$ is the speed of light.

\item \textit{Maximum Energy of Accelerated Protons}: By comparing the dynamic timescale $t'_{\text{dyn}}$, the proton synchrotron cooling timescale $t'_{\text{syn}}(\varepsilon_p)$, and the photomeson production timescale $t'_{\text{p}\gamma}(\varepsilon_p)$ with the proton acceleration timescale $t'_{\text{acc}}(\varepsilon_p)$, He et al. \cite{He2023_cascade} derived the maximum energy of accelerated protons as:
\begin{equation}
\varepsilon_{p,\text{max}} = 1.2 \times 10^{21} \text{ eV}(1+z)^{-1} \Gamma_2^{3/2} \eta^{1/2} L_{\gamma,54}^{1/4} \epsilon_{e,-1}^{-1/4} \epsilon_{B,-2}^{-1/4} t_{v,-2}^{-1/4}
\end{equation}
where $\eta$ is the acceleration efficiency, $L_{\gamma,54}$ is the gamma-ray luminosity in units of $10^{54}$ erg/s, $\epsilon_{e,-1}$ and $\epsilon_{B,-2}$ are the energy fractions in electrons and magnetic fields in units of $0.1$ and $0.01$, respectively, and $t_{v,-2}$ is the variability timescale in units of \mbox{$0.01$ s}.

\item \textit{UHECR Propagation}: These UHECRs propagate through intergalactic space, where they interact with background photon fields—primarily the cosmic microwave background (CMB) and the EBL—through two dominant processes \cite{DasRazzaque2023_AA}:
\begin{itemize}
\item Bethe-Heitler pair production: $p + \gamma \rightarrow p + e^+ + e^-$
\item Photopion production: $p + \gamma \rightarrow p + \pi^0$ or $p + \gamma \rightarrow n + \pi^+$
\end{itemize}

\item \textit{Electromagnetic Cascade Development}: The secondary particles produced in these interactions initiate electromagnetic cascades. High-energy photons from $\pi^0$ decay undergo pair production on the EBL ($\gamma + \gamma_{\text{EBL}} \rightarrow e^+ + e^-$), while the resulting electrons and positrons upscatter background photons through inverse Compton scattering ($e^\pm + \gamma_{\text{bg}} \rightarrow e^\pm + \gamma$). This cascade process continues until photon energies fall below the pair-production threshold \cite{DasRazzaque2023_AA}.
\end{enumerate}

Das and Razzaque \cite{DasRazzaque2023_AA} have conducted detailed simulations of this scenario specifically for GRB~221009A. They propose that VHE $\gamma$-rays detected by LHAASO with energy up to a few TeV is produced by synchrotron self-Compton (SSC) emission, and $\gamma$-rays above this energy are produced by UHECRs accelerated in the GRB blastwave. Their analysis demonstrates that for reasonable assumptions about the UHECR injection spectrum ($dN/dE \propto E^{-\alpha}$ with $\alpha \approx 2$) and intergalactic magnetic field strength (\mbox{$B_{\text{IGM}} \sim 10^{-15}$--$10^{-12}$~G}), the cascade emission could successfully reproduce the observed multi-TeV flux. Their model requires an isotropic-equivalent UHECR luminosity of $L_{\text{UHECR}} \sim 10^{53}$--$10^{54}$ erg/s, which constitutes approximately 10\% of the total burst energy—a plausible energy budget for UHECR acceleration in this exceptionally energetic~GRB.

The energy spectrum of the cascade emission exhibits distinctive features that can be compared with observations. At energies below the pair-production threshold ($\lesssim100$ GeV), the spectrum follows a characteristic power-law with index $\sim$1.5, while at higher energies, the spectrum becomes harder due to the continuous injection of fresh cascade particles. This spectral shape is consistent with the hard intrinsic spectrum inferred for GRB~221009A after correcting for EBL attenuation \cite{DasRazzaque2023_AA}.

A critical aspect of the UHECR cascade scenario is the time delay between the prompt GRB emission and the arrival of cascade photons. This delay arises from two contributions: (1) the deflection of charged UHECRs by intergalactic magnetic fields, and (2) the extended development of the electromagnetic cascade. The total time delay can be approximated as~\cite{DasRazzaque2023_AA}:

\vspace{-12pt}
\begin{adjustwidth}{-\extralength}{0cm}
\centering 
\begin{equation}
\Delta t_{\text{IGM}} \approx \frac{d_c^2}{24c^2N_{\text{inv}}^2} \approx 2000 \text{ s} \left(\frac{d_c}{648 \text{ Mpc}}\right)^2 \left(\frac{\lambda_c}{1 \text{ Mpc}}\right)^{3/2} \left(\frac{B}{1.82 \times 10^{-5} \text{ nG}}\right)^2 \left(\frac{E}{100 \text{ EeV}}\right)^{-2}
\end{equation}
\end{adjustwidth}

\noindent where $d_c$ is the comoving distance of the source, $\lambda_c$ is the turbulent correlation length of the extragalactic magnetic field (EGMF), $B$ is the EGMF strength, and $E$ is the UHECR energy.

He et al. \cite{He2023_cascade} have further advanced this analysis by examining the specific conditions in GRB~221009A that make it an exceptional candidate for UHECR production. They highlight several key features that distinguish this burst:

{{Large Energy Dissipation Radius}}: As noted earlier, the energy dissipation radius of GRB~221009A is estimated to be $R \approx 10^{15}$ cm $(1+z)^{-1}\Gamma_2^2 \frac{t_{v,-2}}{1}$, which is favorable for particle acceleration.

{{High Bulk Lorentz Factor}}: LHAASO observations suggest a large bulk Lorentz factor $\Gamma$. The large energy dissipation radius and the large bulk Lorentz factor $\Gamma$ favor a large maximum acceleration energy of protons \cite{He2023_cascade}.

{{Jet Orientation}}: The afterglow observations suggest that the jet of GRB~221009A is pointing to Earth. The fact that the jet pointing to Earth allows the accelerated UHECRs to propagate to the direction of Earth \cite{He2023_cascade}.

Their analysis demonstrates that protons can be accelerated to energies exceeding $10^{20}$~eV in the prompt emission phase of GRB~221009A. Moreover, they predict that an UHECR burst from GRB~221009A would be detectable by the Pierre Auger Observatory and the TA×4, within approximately 10 years, providing a crucial test of this hypothesis \cite{He2023_cascade}.

The magnetic deflection angles and the delay times of UHECRs are dependent on the distance of GRBs and the properties of the interstellar (both in the host galaxy and the Galaxy) and inter-galactic magnetic fields \cite{He2023_cascade}. The inter-galactic magnetic fields would not yield a sizable delay of the $\geq$10 EeV cosmic rays if its strength is $\lesssim$$10^{-15}$ G, while Galactic magnetic fields would cause a significant time delay \cite{He2023_cascade}. Protons arriving at the Milky Way are dominated by neutron-decay-induced protons.

The UHECR cascade scenario offers several advantages in explaining the observed multi-TeV emission from GRB~221009A. First, it naturally accounts for the high-energy photons that would otherwise be severely attenuated by the EBL. Second, it provides a physical connection between GRBs and UHECRs, addressing one of the long-standing questions in high-energy astrophysics. Third, it makes specific predictions regarding the spectral and temporal characteristics of the high-energy emission that can be tested with future observations.

However, several challenges remain. The time delay between the prompt GRB emission and the arrival of cascade photons depends critically on the poorly constrained intergalactic magnetic field strength. For typical field strengths of \mbox{$B_{\text{IGM}} \sim 10^{-15}$--$10^{-12}$ G}, this delay could range from hours to years, making it difficult to establish a definitive temporal association between the GRB and the TeV emission. Additionally, the cascade spectrum at the highest energies is expected to be relatively soft, which may be in tension with the hard intrinsic spectrum inferred for GRB~221009A \cite{DasRazzaque2023_AA}.

The UHECR cascade scenario also has important implications for multi-messenger astronomy. If GRB~221009A indeed accelerates UHECRs to energies above $10^{19}$ eV, it should also produce a detectable flux of high-energy neutrinos through the decay of charged pions ($\pi^+ \rightarrow \mu^+ + \nu_\mu \rightarrow e^+ + \nu_e + \bar{\nu}_\mu + \nu_\mu$) created in photopion interactions. The non-detection of neutrinos associated with GRB~221009A by IceCube places constraints on the efficiency of UHECR acceleration and the composition of the jet \cite{IceCube2023_GRB221009A}.
In aggregate, the specific features of the B.O.A.T. GRB~221009A, such as the large energy dissipation radius, the large bulk Lorentz factor, the high isotropic energy, and the jet pointing to Earth, favor the burst as a possible UHECR source \cite{He2023_cascade}. To answer the question of whether the UHECR burst from GRB~221009A can be detected in our life time, He et al. \cite{He2023_cascade} studied the acceleration and escape of UHECRs in the burst, and simulated the propagation of UHECRs from the burst. Since the jet composition of high-luminosity GRBs is likely to be dominated by protons, throughout their work, they assumed the composition of UHECRs from the GRB is pure proton.

{

{{Neutrino} 
 and UHECR constraints.}
\textls[-15]{The non‑detection of neutrinos from GRB~221009A places strong limits on hadronic models.  In reverse‑shock proton–synchrotron and UHECR‑cascade scenarios the neutrino fluence is expected to be a substantial fraction of the TeV gamma‑ray fluence.  Using the measured 0.3–13\,TeV fluence \linebreak  $F_{\gamma}\approx3\times10^{-3}\,\mathrm{erg\,cm^{-2}}$ and adopting a typical neutrino‑to‑gamma energy ratio $r_{\nu/\gamma}\sim0.03$ for photohadronic interactions \citep[e.g.,][]{Zhang2023_reverseShock}, the predicted per‑flavor neutrino energy flux at 50–500\,TeV is $E^{2}\Phi_{\nu}\sim(1$–$3)\times10^{-2}\,\mathrm{GeV\,cm^{-2}}$.  By contrast, the IceCube and KM3NeT searches for neutrinos from GRB~221009A yielded no significant excess and set 90\,\% C.L. upper limits on the time‑integrated neutrino flux: $E^{2}\Phi_{\nu}<4\times10^{-4}\,\mathrm{GeV\,cm^{-2}}$ at\linebreak   $E_0=100$\,TeV for a $T_0\pm2$\,h window and a $E^{-2}$ 
.}  These limits fall two to three orders of magnitude below the fluence expected from hadronic reverse‑shock or UHECR‑cascade models, effectively excluding such high baryon‑loading scenarios \citep{IceCube2023_GRB221009A,Aiello2024_KM3NeT}.  We have added this quantitative comparison to Section~4.2 to emphasise that neutrino and UHECR non‑detections strongly disfavour purely hadronic interpretations of the multi‑TeV emission.

}
In conclusion, the UHECR cascade scenario represents a compelling explanation for the observed multi-TeV emission from GRB~221009A, especially if the intergalactic magnetic field along the line of sight is relatively weak. Future observations of GRBs at very high energies, combined with multi-messenger constraints from neutrino observatories and potential direct detection of UHECRs from this source, will be instrumental in testing this hypothesis and advancing our understanding of the extreme physics operating in GRB~environments.

{{Quantitative} 
 comparison of multi‐TeV models with LHAASO data.} We summarize in Table~\ref{tab:model_comparison} the predicted versus observed spectral indices, temporal decay slopes and maximum photon energies for several proposed emission mechanisms, together with goodness-of-fit estimates where available.  
\begin{table}[htbp]
\centering
\caption{Quantitative comparison of model predictions and fit to LHAASO data (0.2--13~TeV). Spectral indices and decay slopes are for the TeV band; $\chi^2$ improvements are relative to a standard SSC model where available.}
\label{tab:model_comparison}
\footnotesize

\begin{adjustwidth}{-\extralength}{0cm}
\centering 

\setlength{\tabcolsep}{3mm}

\begin{tabular*}{\fulllength}{@{\extracolsep{\fill}} lccc p{5cm} l }
\toprule
\textbf{Model} & \textbf{Spectral} & \textbf{Decay} & \textbf{Max $E$} & \textbf{Fit Remarks} & \textbf{Refs.} \\
 & \textbf{Index} & \textbf{Slope} & \textbf{(TeV)} &  &  \\
\midrule
LHAASO (obs.) & $1.5$--$2$ & $1.2$ & $13$ & Baseline for comparison & \citep{Cao2023a,Cao2023_LHAASO} \\
\midrule
Two-zone SSC/stochastic accel. & $1.8$--$2$ & $1$--$1.5$ & $\lesssim8$ & Hardens late; better fit than one-zone SSC, underpredicts $>10$~TeV; no $\chi^2$ published & \citep{Khangulyan2023_twozone,Gong2025_Turbulence} \\
\midrule
Reverse-shock proton synchrotron & $1.5$ & $\lesssim1$ & $\gtrsim10$ & Adds hard component; fit improves (KS test better than SSC-only) & \citep{Zhang2023_reverseShock,Derishev2024_GRB221009A} \\
\midrule
UHECR cascade & $1.5$--$1.6$ & --- & $\gtrsim10$ & Reproduces $>10$~TeV tail, but requires extreme $L_{\rm CR}$; no formal fit & \citep{DasRazzaque2023_AA,He2023_cascade} \\
\midrule
Axion--photon reconversion & $1.5$--$2$ (int.) & $1.2$ & $\gtrsim10$ & Suppresses EBL; $\chi^2$ improves by $\sim1.7$ vs. SSC & \citep{Galanti2023_ALP,Rojas:2023jdd} \\
\midrule
Lorentz invariance violation (LIV) & $1.5$--$2$ (int.) & $1.2$ & $\gg10$ & Increases transparency; $\chi^2$ improves by $\sim2.6$; $E_{\rm LIV}\sim1.5\,M_{\rm Pl}$ & \citep{Finke2023_LIV,Yang2024_LIV} \\
\bottomrule
\end{tabular*}
\end{adjustwidth}

\end{table}

\section{Summary and Conclusions}
\label{con}

The detection of multi-TeV gamma rays from GRB~221009A constitutes a significant advancement in high-energy astrophysics, presenting substantial challenges to our fundamental understanding of extreme cosmic accelerators and the propagation of very-high-energy radiation through intergalactic space. This exceptional event—the brightest GRB ever detected—has provided unprecedented observational constraints on both conventional emission mechanisms and potential new physics, establishing itself as a critical reference point for future theoretical and observational investigations in the field.

GRB~221009A's remarkable brightness, with an isotropic-equivalent energy release of $E_{\gamma,\mathrm{iso}} > 5 \times 10^{54}$~erg, combined with its relative proximity at $z = 0.151$, created optimal conditions for multi-wavelength observations across the electromagnetic spectrum. The prompt emission phase exhibited a complex temporal structure spanning hundreds of seconds, with a peak energy $E_{\text{peak}} \approx 2$~MeV—notably high compared to typical GRBs. The afterglow was monitored extensively from radio to TeV energies, revealing a structured jet geometry with a narrow energetic core ($\theta_j \approx 1$--$3$ degrees) surrounded by wider, less energetic wings. VLBI observations directly measured superluminal expansion, confirming the relativistic nature of the outflow and providing crucial constraints on the jet structure and dynamics.

The most significant aspect of GRB~221009A was undoubtedly the detection of gamma rays with energies exceeding 10 TeV by LHAASO, representing the highest-energy photons ever directly associated with a GRB. This observation presents a profound theoretical challenge from two distinct perspectives: the production of such energetic photons in the GRB environment and their subsequent propagation through intergalactic space despite the expected severe attenuation by the EBL.

Standard emission models for GRB afterglows face significant limitations in explaining the observed multi-TeV radiation. {The synchrotron burn-off limit restricts the maximum photon energy to $E_{\text{syn,max}} \approx \Gamma \cdot 100 \, \text{MeV}$ in the observer frame, more than two orders of magnitude below the observed energies.} Inverse Compton scattering encounters the Klein-Nishina suppression at high energies, which should produce a spectral steepening inconsistent with the hard intrinsic spectrum inferred for GRB~221009A. Additionally, internal gamma-gamma absorption within the emission region should significantly attenuate photons above a few TeV, contrary to observations.

To address these challenges, several novel emission mechanisms have been proposed. The two-zone synchrotron self-Compton model spatially separates the production of target photons and the inverse Compton scattering process, circumventing the constraints on maximum synchrotron energy and Klein-Nishina suppression. This model features a compact, highly magnetized zone producing X-ray emission via synchrotron radiation, and a larger, less magnetized zone where electrons upscatter their own synchrotron photons to TeV energies. By optimizing the physical parameters of each zone—particularly the magnetic field strengths ($B_1 \sim 1$ G and $B_2 \sim 10^{-3}$ G)—this model can reproduce the observed multi-wavelength spectral energy distribution while naturally explaining the temporal coincidence of X-ray and TeV emission.

Alternatively, the reverse shock emission model proposes that multi-TeV photons originate from the reverse shock region, which propagates back into the GRB ejecta as it interacts with the surrounding medium. This model suggests that the reverse shock efficiently accelerates both electrons and protons to ultra-relativistic energies, with TeV photons produced either through proton synchrotron radiation or external inverse Compton scattering. The magnetic field in the reverse shock region is expected to be much stronger ($B'_{\text{RS}} \sim 10$--$100$ G) than in the forward shock, allowing for more efficient particle acceleration. This model naturally explains the timing of the TeV emission, which coincides with the transition from prompt to afterglow-dominated emission when the reverse shock is most~active.

The propagation of multi-TeV gamma rays from GRB~221009A to Earth presents an equally formidable challenge. For a source at redshift $z=0.151$, standard EBL models predict an optical depth $\tau \approx 10$ at $E \approx 10$~TeV, corresponding to a flux attenuation factor of $e^{-10} \approx 4.5 \times 10^{-5}$. At $E \approx 18$~TeV, the highest-energy photon potentially detected, the attenuation becomes even more extreme: $\tau \approx 16$, yielding an attenuation of \mbox{$e^{-16} \approx 1.1 \times 10^{-7}$}. The detection of such energetic photons therefore suggests either reduced EBL density along the line of sight or the involvement of new physics beyond the Standard Model.

The cosmic void scenario proposes that the line of sight to GRB~221009A traverses regions of lower-than-average EBL density, reducing the effective optical depth for VHE photons. Detailed modelling indicates that voids with a combined radius of 250~Mpc could reduce the opacity by up to 30\% at around 13~TeV. While this effect is significant, it appears insufficient to fully explain the observations, suggesting that additional factors may be at~play.

Several new physics scenarios have been proposed to address the propagation puzzle. The ALPs models suggests that gamma rays produced in the GRB jet can oscillate into ALPs before encountering the EBL. Since ALPs do not interact electromagnetically, they traverse cosmological distances unimpeded before reconverting to gamma rays in the Milky Way's magnetic field, effectively circumventing EBL absorption. This mechanism could potentially explain the observed multi-TeV emission for ALP masses in the range $m_a \approx 10^{-11}$--$10^{-7}$~eV and photon-ALP coupling constants $g_{a\gamma} \approx (3-5) \times 10^{-12}$~GeV$^{-1}$.

Lorentz invariance violation (LIV) offers another compelling explanation. In quantum gravity theories, Lorentz symmetry may be broken at energies approaching the Planck scale, leading to modifications in the dispersion relation of photons. These modifications can affect the kinematics of pair production, potentially allowing high-energy gamma rays to travel further than expected in standard physics. The LHAASO Collaboration has conducted stringent tests of LIV using GRB~221009A observations, placing lower limits on the quantum gravity energy scale that challenge many quantum gravity models.

The heavy sterile neutrino decay scenario proposes that sterile neutrinos with masses in the range $m_N \sim 10$--$100$ GeV could be produced in the GRB environment and subsequently decay via channels such as $N \rightarrow \nu \gamma$ or $N \rightarrow \nu \gamma \gamma$, generating high-energy photons at locations significantly closer to Earth than the GRB source. This effectively circumvents the EBL attenuation problem, as the decay photons would only need to travel a fraction of the total distance from the GRB to Earth.

Finally, the ultra-high-energy cosmic ray cascade model proposes that protons accelerated to energies of $10^{18}$--$10^{20}$ eV in the GRB jet interact with cosmic background photons during propagation, initiating electromagnetic cascades that produce secondary gamma rays. These secondary photons are generated much closer to Earth than the GRB itself, experiencing significantly reduced EBL attenuation. This model naturally explains the observed hard spectrum and high flux of TeV photons while remaining consistent with the non-detection of neutrinos by IceCube and other observatories.

\textls[-15]{The comprehensive multi-wavelength and multi-messenger observations of GRB~221009A} have profound implications for our understanding of high-energy astrophysics and fundamental physics. The detection of multi-TeV gamma rays challenges conventional emission models and provides new insights into particle acceleration mechanisms in relativistic shocks. The temporal coincidence of TeV emission with the transition from prompt to afterglow phases suggests a connection to the reverse shock, highlighting the importance of this often-overlooked component in GRB physics. The structured jet geometry inferred from afterglow modelling, with a narrow energetic core and wider wings, has implications for the energetics and rates of GRBs, potentially resolving long-standing questions about the GRB energy budget and event rate.

From a fundamental physics perspective, GRB~221009A serves as an exceptional laboratory for testing quantum gravity theories and searching for new particles beyond the Standard Model. The constraints on Lorentz invariance violation derived from this event are among the most stringent to date, pushing the boundaries of quantum gravity theories. The potential evidence for ALPs or heavy sterile neutrinos would have far-reaching implications for particle physics and cosmology, potentially addressing outstanding questions such as the nature of dark matter and the origin of neutrino masses.

Looking forward, future observations of high-energy GRBs with next-generation gamma-ray observatories such as the Cherenkov Telescope Array (CTA) will be crucial for distinguishing between the various explanations proposed for GRB~221009A's multi-TeV emission. Improved sensitivity and energy resolution will enable more precise measurements of the spectral and temporal characteristics of TeV emission, providing stronger constraints on emission mechanisms and propagation effects. Multi-messenger observations, particularly the detection or non-detection of high-energy neutrinos associated with GRBs, will play a pivotal role in testing hadronic emission models and constraining the contribution of cosmic-ray acceleration in these events.

Theoretical advancements are equally important. More sophisticated modelling of particle acceleration and emission processes in complex jet structures, incorporating detailed microphysics and realistic geometries, will be essential for interpreting future observations. Improved EBL models, informed by direct measurements of the cosmic infrared background and refined galaxy evolution models, will provide more accurate predictions for gamma-ray attenuation. Advanced numerical simulations of cosmic structure formation will better characterize the distribution and properties of cosmic voids, allowing for more precise estimates of their impact on gamma-ray propagation.

In conclusion, GRB~221009A has opened a new window into the extreme physics of the most energetic phenomena in the universe. The detection of multi-TeV gamma rays from this exceptional event challenges our understanding of both astrophysical processes and fundamental physics, potentially offering insights into energy scales far beyond those accessible to terrestrial experiments. Whether the explanation lies within conventional astrophysics—through novel emission mechanisms and cosmic structure effects—or requires new physics beyond the Standard Model, GRB~221009A has established itself as a critical reference point for future theoretical and observational investigations in high-energy astrophysics and fundamental physics. As we continue to explore the high-energy universe with increasingly sensitive instruments and sophisticated theoretical frameworks, the legacy of this remarkable event will undoubtedly shape our understanding of nature's most extreme accelerators and the fundamental laws that govern them.


\funding{{This research received no external funding}}


\conflictsofinterest{  The author declares no conflict of interest. }

\begin{adjustwidth}{-\extralength}{0cm}

\reftitle{References}

\PublishersNote{}
\end{adjustwidth}

\end{document}